\documentclass[9pt,a4paper,oneside]{article} 
\usepackage[numbers,sort&compress]{natbib}

\usepackage[paper=a4paper,left=20mm,right=20mm,top=25mm,bottom=25mm]{geometry}
\usepackage{graphicx}
\usepackage[titletoc]{appendix}
\usepackage{amsmath}
\usepackage{amssymb}
\usepackage{xcolor}
\usepackage{empheq}
\definecolor{TURed}{RGB}{153,0,0}
\definecolor{DarkGreen}{RGB}{0,153,0}
\definecolor{DarkBlue}{RGB}{0,0,153}
\newcommand{\red}[1]{\textcolor{black}{{#1}}}
\newcommand{\blue}[1]{\textcolor{black}{#1}}

\newcommand{\white}[1]{\textcolor{white}{#1}}

\newcommand{\nn}{\nonumber\\}
\newcommand{\xt}{x_\tau}
\newcommand{\xtt}{x_{2\tau}}

\begin{document}

\title{\textbf{Fokker-Planck equations for time-delayed systems\\via Markovian Embedding}}%

\author{Sarah A. M. Loos\thanks{sarahloos@itp.tu-berlin.de}         \and
        Sabine H. L. Klapp 
}
\date{  \small     Institut f\"ur Theoretische Physik,
        Technische Universit\"at Berlin,\\
                        Hardenbergstr.~36,
                        D-10623 Berlin,
                        Germany
}

\maketitle

\begin{abstract} 
For stochastic systems with discrete time delay, the Fokker-Planck equation (FPE) of the one-time probability density function (PDF) does not provide a complete, self-contained probabilistic description. It explicitly involves the two-time PDF\red{, and represents, in fact, only the first member of an infinite hierarchy}. We here introduce a new approach to obtain a Fokker-Planck description by using a Markovian embedding technique and a subsequent limiting procedure. On this way, we {find a closed, complete FPE in an infinite-dimensional space, from which one can} derive a hierarchy of FPEs. While the first member is the well-known FPE for the one-time PDF, we obtain, as second member, a new representation of the equation for the two-time PDF. From a conceptual point of view, our approach \red{is simpler than earlier derivations and it} yields interesting insight into both, the physical meaning, and the mathematical structure of delayed processes. \red{We further propose an approximation for the two-time PDF,
which is a central quantity in the description of these non-Markovian systems as it directly gives the correlation between the present and the delayed state. Application to a prototypical bistable system reveals that this approximation captures the non-trivial effects induced by the delay remarkably well, despite its surprisingly simple form. Moreover, it outperforms earlier approaches for the one-time PDF in the regime of large delays.}
\end{abstract}
\setcounter{tocdepth}{5}
%
\section{Introduction}
\label{intro}
Predicting stochastic dynamics far from thermal equilibrium is a major goal of statistical physics and a key step towards an understanding of realistic, fluctuation-dominated
systems from biology, chemistry, or socio-economics, just to name a few of the many fields of application.
 An important mechanism that drives system out of equilibrium and, at the same time, renders the dynamics non-Markovian is \textit{time delay}. Delays can have various origins, such as unavoidable signal transmission time lags (e.\,g., in laser systems \cite{Kane2005,Liu2001}), information processing times (as in neural systems~\cite{Waibel1995,Longtin1990,Cabral2014,Gupta2011}), or decision making times (e.\,g., in the context of financial markets \cite{Tambue2015,Callen2013}). Moreover, they are omnipresent in systems subject to feedback control~\cite{Schoell2016,Carmele2013,Nemet2016,Durve2018,Khadka2018,Loos2014,Mijalkov2016,Bruot2011,Masoller2002,Gernert2016}.
In many cases, it is sufficient and suitable to model such processes by including a single, \textit{discrete} delay and a white noise. This is the situation considered in this paper. Discrete delays are known to induce complex behavior (e.\,g., delay-induced oscillations, chaos), and, on the other hand, to stabilize unstable orbits~\cite{Schneider2013,Zakharova2016,Schneider2016}. Therefore, they are often implemented intentionally, for example in the framework of Pyragas control~\cite{Schoell2008}. 
{It is important to} note that the time-delayed systems considered here are different from those modeled by generalized Langevin equations~\cite{Zwanzig1973}. The latter involve a memory kernel in the deterministic (friction) force resulting from conservative interactions between a colloidal particle (or, more generally, a ``slow'' variable) and the bath particles (as well as among the bath particles). This delayed friction is related to the noise via a fluctuation-dissipation relation~\cite{Kubo1966} (see, e.\,g.,~\cite{Krueger2016,Bao2005,Maes2013,Maes2014}). In contrast, the systems considered here involve a discrete delay within a deterministic force, and there is no relation to noise correlations.
  \par
While they have numerous applications, the mathematical description of time-delayed stochastic systems is still far from being complete. The Langevin equation~(LE) is used to model such processes on a stochastic level, but it is infinite-dimensional due to the discrete delay. Exact analytical results based on the LE have, so far, been obtained only for linear systems~\cite{Kuechler1992}. On the other hand, the Fokker-Planck (FP) description for the probability density function (PDF), {which we focus on in this paper, is}
given by an \textit{infinite hierarchy} of coupled equations of increasing complexity~\cite{Frank2005a,Loos2017,Rosinberg2015,Giuggioli2016}. As shown by Frank~\cite{Frank2005a} (who only considered the first member), this hierarchy can be derived using \textit{Novikov's theorem}~\cite{Novikov1965}. The first member of the FP hierarchy (originally derived by Guillouzic et al.~\cite{Guillouzic1999}) is the FP equation (FPE) for the one-time PDF. It explicitly involves the two-time PDF for the spatio-temporal correlations between times $t$ and $t-\tau$, with $\tau$ being the delay time. Although not being self-sufficient, the equation for the one-time PDF has proven itself to be quite useful for the analytical treatment. On the one hand, it is an important tool in the search for exact results~\cite{Frank2001,Loos2019}. Moreover, on the base of this FPE different approximation schemes have been developed in the past years (see~\cite{Loos2017} for an overview). Examples are the perturbation theory~\cite{Frank2005} and the ``force-linearization closure''~\cite{Loos2017,Loos2019}, both of which yield reasonable approximations for the one-time PDFs in quite large parameter regimes. (As opposed to this, the so called small $\tau$ expansion~\cite{Guillouzic1999}, which is based on the LE, is only appropriate for very small values of the delay time, where the system behaves more or less like a Markovian one.) \par
While these analytical results have indeed contributed to an understanding of delayed processes, major challenges remain. In particular, the insights provided by
the {one}-time PDF \textit{alone} are admittedly limited as we deal with non-Markovian dynamics, which is crucially determined by memory, and, hence, temporal \textit{correlations}. Having this in mind, it is not surprising that many non-Markovian effects are only captured by two-time PDFs (or higher PDFs). In fact, only by studying them, non-Markovian steady-states can be qualitatively distinguished from thermal equilibria at all\footnote{As an illustrative example, consider the following. Let $\rho^\mathrm{nM}_1(x)$ be the non-Markovian one-time PDF of an arbitrary nonequilibrium steady state. This PDF can in general not be distinguished from a PDF $\rho^\mathrm{M}_1(x)$ of a Markovian (equilibrium) process in a (fictive) potential defined by $U_\mathrm{ficitve}(x)\equiv k_\mathrm{B}\mathcal{T} \ln{\rho^\mathrm{nM}_1(x)}$, which would equally yield a distribution $\rho^\mathrm{M}_1 =Z^{-1}e^{-U_\mathrm{fictive}/k_\mathrm{B}\mathcal{T}}=\rho^\mathrm{nM}_1$.}.
{Moreover, the two-time PDF is essential to obtain stochastic thermodynamical quantities like the fluctuating heat, work, or entropy~\cite{Seifert2012}. As an example, the $m^\mathrm{th}$ moment of the heat rate~\cite{Sekimoto2010,Loos2019} of a system described by the LE~(\ref{EQ:LE-orignial}) is given by $\langle \dot{q}(t)^m\rangle =\langle F[X(t),X(t-\tau)]^{2m}\rangle+const.$, which means that all moments can be evaluated via the two-time PDF. Furthermore, calculations of transport properties such as, e.\,g., the Kramers rates, that are solely based on the one-time PDF have been shown to be limited to parameter regimes where non-Markovian effects are rather weak~\cite{Loos2017}. However, despite the clear benefits of a probabilistic treatment and the apparent limits of a study focusing on the first member of the FP hierarchy alone, no general approximation schemes for the two-time PDF or other temporal correlation functions are known (one exception being the two-state reduction employed in~\cite{Tsimring2001,Loos2019}). In fact, the higher FP hierarchy members have rarely been discussed at all. This hinders the development of further approximation schemes and solution methods. 
In the present work we concentrate on these higher members. 
By introducing a different FP approach which yields an alternative representation of the hierarchy, we provide a step towards a full understanding of the probabilistic description. \par
Contrary to earlier approaches, we here derive a FP hierarchy by using a Markovian embedding technique~\cite{Puglisi2009,Crisanti2012,Siegle2010,Siegle2011,Bao2005,Villamaina2009}, which is inspired by the \textit{linear chain trick}~\cite{Longtin2010,Smith2011,Niculescu2012}. In particular, we introduce $n\in \mathbb{N}$ auxiliary variables whose dynamics is given by a set of $n$ coupled, linear, stochastic differential equations without delay. This set generates the same stochastic process as the delayed equation in the limit $n\to \infty$. The resulting Markovian system is described by a closed, \textit{self-sufficient} FPE on an $n$-dimensional space. By marginalization, i.\,e., projection onto a one-dimensional subspace, and subsequent limiting procedure, we obtain the well-known delayed FPE for the one-time PDF. Furthermore, by projection onto higher-dimensional subspaces, we derive a FP hierarchy which is different from the one obtained by Novikov's theorem~\cite{Novikov1965}. \red{Interestingly, the limiting procedure contains one step where position-like variables are converted into velocity-like variables.} Explicitly considering the second member, we show how the two hierarchies can be converted into each other. 
\par
\red{Furthermore, we propose an approximation for the steady-state probability density in the regime of large delay times, based on the new equation for the two-time PDF. More specifically, by neglecting correlations between the system state and the displacement, we obtain a closed equation that can be solved analytically. To test the approximation, we consider a bistable system involving a doublewell potential and a linear delay force. This is an established prototypical model to study the combined impact of noise and time delay on nonlinear dynamics~\cite{Tsimring2001,Loos2019,Loos2017}. In particular, the model exhibits delay-induced particle oscillations between the two minima. As these oscillations manifest in the two-time PDF, this system is an ideal candidate to test approximations, and to study non-Markovian effects on the level of the two-time PDF.
By comparing with numerical simulations, we demonstrate that the proposed approximation yields surprisingly good results. In particular, the approximation captures the main characteristics of the two-time PDF, both, when the delay force is directed towards, and when it points away from the delayed position. In this way, we present to our knowledge the first approximation for the two-time probability density of stochastic systems with time delay. In the regime of large delay times, the new approximation even outperforms earlier approaches known from the literature, demonstrating that a framework on the level of the second FP member, is somewhat superior compared to approaches on the first member.}
\par
This paper is organized as follows. First, we review earlier derivations of the FPE known from the literature. Then, we discuss the Markovian embedding and show how the delayed FPE can be derived from it by marginalization. We further discuss the higher members stemming from this approach, particularly focusing on the second member, which are different from those obtained by earlier approaches.
\red{At the end of the main text, we introduce the aforementioned approximation scheme and discuss the application to the bistable, delayed system.}
 In the appendix, we show how the second member of our hierarchy can be transformed into the one obtained from Novikov's theorem, and present a short derivation of Novikov's theorem via path integrals. 
In this work, we restrict ourselves to systems in one spatial dimension with a single time delay. However, our results can be readily generalized to multidimensional systems, and multiple delays can, as well, be treated analogously.
%
\section{Langevin equation and Fokker-Planck equation for delayed systems}
We consider a system described by a stochastic delay differential equation, in particular, by the overdamped LE
\begin{equation}  \label{EQ:LE-orignial}
\mathrm{d}{X}_{}(t)= {F_{}}[X_{}(t),{X_{}(t-\tau)}]\mathrm{d}t+\textcolor{black}{\sqrt{2 D_0}\,\mathrm{d}W(t) },
\end{equation}
with $X\in \Omega=[\omega_1,\omega_2] \subseteq \mathbb{R}$ and time $t >0$. $F_{}$ denotes the deterministic force\footnote{The force is scaled with the friction coefficient $\gamma$, i.\,e., $F \to F/\gamma$, which is related with the bath's temperature $\mathcal{T}$ via $\gamma D_0 = k_\mathrm{B}\mathcal{T}$, where $k_\mathrm{B}$ is the Boltzmann constant.} given by some (generally nonlinear) function depending on the instantaneous and on the delayed system state, $X(t)$ and $X(t-\tau)$, with  $\tau\geq0$ being the single discrete {time delay}. $W$ is a Wiener process with independent increments, $\langle W \rangle =0$, and $\langle W(t)W(t') \rangle = \min(t,t')$, generated by the additive Gaussian white noise $\xi$, with $\langle \xi \rangle =0$, $\langle \xi(t)\xi(t') \rangle = \delta(t-t')$. Thereby, $\langle ... \rangle$ denotes the ensemble average (i.\,e., average over all noise realizations), w.\,r.\,t. a given initial condition. The latter is specified by a history function $\phi(\tau \leq t\leq 0)$, i.\,e., $X(t)=\phi(t), t\in [-\tau,0]$, which can be fixed for the entire ensemble or be drawn from a distribution $P(\phi)$. $D_0$ is the (constant) strength of the noise. We further consider {natural boundary conditions} (details are given below).

The FP description provides a complementary way to model the process on the probabilistic level. It is well-known that the FPE for the one-time PDF related to (\ref{EQ:LE-orignial}) reads~\cite{Frank2004,Guillouzic1999,Frank2005a,Loos2017}
\begin{align}\label{EQ:dFPE_start}
{ \partial_t} \rho_1( x ,t) = - \partial_{x } \int_{\Omega}^{ }\left[ F_{}(x,x_\tau) \rho_{2}(x ,t;x_\tau,t-\tau )\right]\mathrm{d}x_\tau +D_0 \, \partial_{x }^2 \rho_1( x,t ),
\end{align}
where $\rho_N(x ,t;\xt,t-\tau;...;x_{N\tau},t-N\tau)=\langle \delta(x =X(t))\delta(x_\tau=X(t-\tau))...\delta(x_{N\tau}=X(t-N\tau)) \rangle$ denotes the $N$-time PDF (with normalization $\iint_{\Omega}...\int_{\Omega}\rho_N \mathrm{d}x\mathrm{d}x_\tau...\mathrm{d}x_{N\tau}=1$).
As opposed to Markovian dynamics, the FPE for $\rho_1$ is not closed but involves the two-time (joint) PDF, $\rho_2$. The FPE~(\ref{EQ:dFPE_start}) can be derived by different strategies~\cite{Guillouzic1999,Frank2005a,Frank2005,Frank2004,Zheng2017}. 
However, only the approach which bases on Novikov's theorem, results in a complete FP description (in the form of an infinite hierarchy, see below), while all the others only yield the first, not self-sufficient member. In the following, we briefly review earlier work, in particular Refs.~\cite{Frank2005a,Frank2004,Zheng2017}, in order to better understand the technical and conceptual difference of our approach.
\subsection{Second member of the Fokker-Planck hierarchy from Novikov's theorem}
In this section, we briefly review the approach~\cite{Frank2005a,Loos2017} based on {Novikov's theorem}~(\ref{NOV-THE}) (a short derivation of the latter is given in the Appendix~\ref{APP:Nov-Theorem}). 
Besides leading to~(\ref{EQ:dFPE_start}), this approach can also be used to obtain the FPE for $\rho_2$ and, in general, $\rho_N, ~N\in\mathbb{N}$. It renders an infinite hierarchy of coupled equations of increasing complexity, whose $N$th member involves both, the $N$-time (joint) PDF $\rho_N(x,t;x_\tau,t-\tau;x_{2\tau}, t-2\tau; .. ;x_{N},t-N\tau)$ and the $(N+1)$-time PDF $\rho_{N+1}$. %
It should be noted, however, that the higher members (starting from the second one) contain functional derivatives %
w.\,r.\,t the noise, which need to be obtained separately from the LE. For example, the second member reads
\begin{align}\label{EQ:General_dFPE2}
{ \partial_t} \rho_2(x,t;x_\tau,t\!-\!\tau)=&   -{\partial_x} \left\lbrace F_{}(x,x_\tau) \rho_2(x,t;x_\tau,t\!-\!\tau) \right\rbrace %
-\partial_{x_\tau} \left\lbrace  \int_{\Omega} F (x_\tau,x_{2\tau}) \rho_3(x,t;x_\tau,t\!-\!\tau;x_{2\tau},t\!-\!{2\tau}) \mathrm{d}x_{2\tau}\!\right\rbrace \nonumber  \\
& + D_0 \,\Bigg( {\partial_{x}^2} + {\partial_{x_{\tau}}^2}  
+{\partial_{x}\partial_{x_{\tau}}} \frac{\delta X(t)}{\delta \xi(t\!-\!\tau)}\Big|_{\!\!\!\!\!\!\! \!\!\!  X(t)=x \atop\! \!  X(t-\tau)=x_\tau}  \Bigg) \rho_2(x,t;x_\tau,t\!-\!\tau) 
\end{align}
(see~\cite{Loos2017} for a derivation).
This equation resembles a Markovian FPE of a two-dimensional (or two-particle) process in the variables $x$ and $\xt$, with the exception of the term which involves a derivative w.\,r.\,t. both variables and the functional derivative w.\,r.\,t. the noise. For the special case of linear forces $F$, the latter can be calculated using the method of steps~\cite{Frank2003}, see Appendix~\ref{APP:Nov-Embedd-Equivalence}. However, to the best of our knowledge, this functional derivative is unknown for general nonlinear forces. We note that this term stems from the fact that if $X(t)$ and $X(t-\tau)$ are regarded as ``two'' random processes, then these processes are not only connected by the force $F_{}$ (via the LE), but additionally by the fact that the corresponding noise processes $\xi(t)$ and $\xi(t-\tau)$ are \textit{identical} with time-shift $\tau$. %
To elucidate this last point, we briefly review an alternative derivation of the FPE known from the literature, which does not yield this term, but instead imposes additional constraints on the solutions.
\par
{The mentioned approach is used in~\cite{Frank2004,Zheng2017} to derive a FPE for $\rho_1$. Here we extended it towards the higher members of the hierarchy. The approach uses a description with \textit{two} time arguments $j$, $z$. In particular, an integer $j=0,1,2,..$ counts the number of intervals of length $\tau$ that have passed since the beginning $t=0$, and a continuous variable $z\in [0,\tau]$ measures the time within the current interval. Then, auxiliary phase-space variables $X_{j \tau}(z) := X(z+j\tau)$ are introduced, which, by construction, all follow a LE of the identical form 
\begin{align}\label{EQ:2timevarapp}
\mathrm{d}{X}_{j\tau}(z)= {F_{}}[X_{j\tau}(z),{X_{(j-1)\tau}(z)}]\mathrm{d}z+{\sqrt{2 D_0}\,\,\mathrm{d}W_{j\tau}(z) }.
\end{align}
(\red{The steady-state dynamics can be accessed if} $j\to \infty$.) A sketch of the construction of the auxiliary variables is given in Fig.~\ref{FIG:1}. In the probability space belonging to the phase space $\{X_{0 \tau}(z),...,X_{N \tau}(z)\}$, the process is Markovian, and the corresponding FPE is given by
\begin{align}\label{EQ:FPE_2-var-approach}
 \partial_{z} \rho_{N+1}
=&\sum_{j=1}^{N} \left[- \partial_{x_{j\tau} } \left\{  F_{}( x_{j\tau}, x_{(j-1)\tau}) \rho_{N+1}\right\} + D_0\partial_{x_{j\tau}}^2  \rho_{N+1} \right]
-\partial_{x_{0\tau}}\left[ F_{}( x_{0\tau}, x_{1\tau})  \rho_{N+1}\right]+ D_0\partial_{x_{0\tau}}^2   \rho_{N+1},
\end{align}
with $\rho_{N+1}(x_{0\tau},x_{1\tau},...,x_{N\tau},z)=\langle \delta[x_{0\tau}-X(z)]\delta[x_{1\tau}-X(z+\tau)]...\delta[x_{N\tau}-X(z+N\tau)] \rangle$. By integrating the equation over the entire domain of all variables but the first (and the last) one, 
this many-variable FPE 
can be used to derive the FPE~(\ref{EQ:dFPE_start}) for $\rho_1$, and a FPE for $\rho_2$, reading
\begin{align}\label{EQ:FPE_2-var-approach_2}
\partial_{t}  \rho_2 =- \partial_{x }  \left[ F_{}(x,x_\tau) \rho_{2}\right]
- \partial_{\xt}  \int_{\Omega}^{ }\left[ F_{}(x,\xt) \rho_{3}\right]\mathrm{d}\xtt
 +D_0 \left[ \partial_{x}^2 + \partial_{\xt}^2 \right]\rho_2,
\end{align}
with $\rho_2(x,t;x_\tau,t-\tau)$ and $\rho_3(x,t;x_\tau,t-\tau;x_{2\tau},t-2\tau)$.
As opposed to the corresponding equation from Novikov's theorem, Eq.~(\ref{EQ:General_dFPE2}), this equation lacks of the term with the unknown functional derivative. 
This is because the stochastic set of Eqs.~(\ref{EQ:2timevarapp}) is, in fact, only equivalent to the delayed process, if one further imposes \textit{additional constraints} 
$X_j(\tau) {=}X_{j+1}(0),~\forall j$, 
ensuring that the end point of the stochastic trajectories on $[j\tau,(j+1)\tau]$ matches the starting point on $[(j+1)\tau,(j+2)\tau]$. Translating these constraints to the level of PDFs is a non-trivial task on its own, thus, the usefulness of these equations as a starting point for approximation schemes is very limited.
The comparison between Eq.~(\ref{EQ:General_dFPE2}) from Novikov's theorem and~(\ref{EQ:FPE_2-var-approach_2}) 
suggests that the terms with the unknown functional derivatives in~(\ref{EQ:General_dFPE2}) are indeed such terms ensuring \red{these} additional constraints on the trajectories. 
\begin{figure}
\includegraphics[width=.52\textwidth]{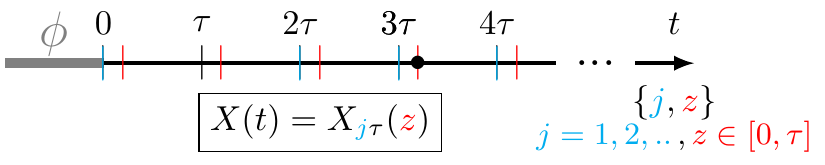}\includegraphics[width=.45\textwidth]{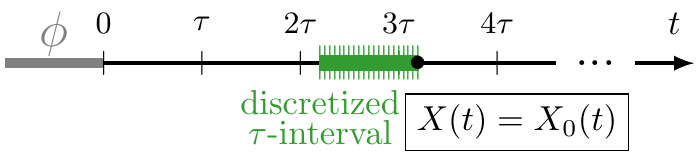}
\caption{Sketch of the different ways to introduce auxiliary variables, \textit{left}: within the approach based on two time arguments~(\ref{EQ:2timevarapp}), \textit{right}: within the Markovian embedding approach~(\ref{EQ:extended-LE}).}\label{FIG:1}
\end{figure}
}
%
\section{Markovian embedding}
As an alternative approach to formulate a complete FP description and to derive the FPE~(\ref{EQ:dFPE_start}), we here employ the Markovian embedding technique~\cite{Puglisi2009,Crisanti2012,Siegle2010,Siegle2011,Bao2005,Villamaina2009}, as also suggested in earlier literature~\cite{Niculescu2012}.
The key idea is to introduce $n\in \mathbb{N}$ auxiliary variables $X_j$ (with one of them replacing the delayed system state in the original equation), giving rise to a $(n+1)$-dimensional Markovian system that generates the same dynamics as the original delay equation (which is hence ``embedded'' in that extended system). Specifically, we consider the set of dynamical equations given by~\cite{Longtin2010}
\begin{subequations} \label{EQ:extended-LE}
\begin{eqnarray}
\dot{X}_0(t)=&F[X_0(t) , X_n(t)]  +{\sqrt{2 D_0 }\,\xi(t) } \label{EQ:Aux-Eq0}\\
\dot{X}_j(t)=&(n/\tau) [X_{j-1}(t)-X_j(t)] , \label{EQ:Aux-Eq1}
\end{eqnarray}
\end{subequations} with $j\in \{1,2,...,n\}$, $t\geq -\tau$, $X_j \in \mathbb{R}$
and the initial conditions $X_{j\in \{1,..,n\}}(-\tau)\equiv0$ (see Appendix~\ref{SEC:App1} for a discussion of the initial conditions).
\textit{Projection} of (\ref{EQ:Aux-Eq1}) onto the variable $X_0$ (see Appendix~\ref{SEC:App1} for details) yields
\begin{equation} \label{EQ:LE-kernel}
\dot{X}_{0}(t)= F\left[ X_{0}(t), \int\limits_{-\tau}^{t} \underbrace{\left( \frac{n}{\tau }\right)^n\frac{(t-s)^{n-1}}{(n-1)!}  \,e^{-n (t-s)/\tau }}_{=K_n}X_{0}(s)\,\mathrm{d}s \right] +\textcolor{black}{\sqrt{2  D_0 }\,\,\xi(t) },
\end{equation}
with the Gamma-distributed memory kernel $K_n$. %
In the limit $n  \rightarrow \infty$, the kernel collapses onto a delta peak at $\tau$, i.\,e., $K_n(t-s)\to\delta(t-s-\tau)$.
Then, the projected Eq.~(\ref{EQ:LE-kernel}) becomes \textit{identical} to the original, delayed LE~(\ref{EQ:LE-orignial})
, with
\begin{align}\label{EQ:Meaning_Aux}
X_0(t)=X(t), ~~~~~~~
X_n(\tau)=X(t-\tau).
\end{align}
In addition to $X_0$ and $X_n$, also the other auxiliary variables can be linked to the variable $X$ of the delayed LE~(\ref{EQ:LE-orignial}). 
In particular, using the difference quotient to approximate $\dot{X}_{j}$ in the dynamical Eqs.~(\ref{EQ:Aux-Eq1}), that is, $[X_{j-1}(t)-X_{j}(t)](n/\tau)=\dot{X}_{j}(t) \approx [X_{j}(t+\Delta t)-X_{j}(t)] /\Delta t  $, gives rise to the interpretations $\Delta t = \tau/n$, and $X_{j}(t+\Delta t)\approx X_{j-1}(t)$. Iteration implies $X_{j}(t+j \tau/n)\approx X_{j-i}(t)$. In combination with Eq.~(\ref{EQ:Meaning_Aux}), this reveals 
\begin{equation}\label{EQ:Interpretation_Xj}
X_{j}(t)= X(t-j \tau/n), ~~\forall j\in \{0,1,2,..,n\}
\end{equation}
which becomes accurate in the limit $n\to\infty$ (at $\Delta t \to 0$ the difference quotient equals the differential quotient).

The here presented scheme of rewriting the delayed differential equation as a set of non-delayed ones is called the \textit{linear chain trick} \cite{Longtin2010,Smith2011,Niculescu2012}. A schematic visualization of the system described by~(\ref{EQ:extended-LE}) is given in Fig.~\ref{FIG:2}.
Importantly, the $n$ auxiliary variables generally follow \textit{linear} Eqs.~(\ref{EQ:Aux-Eq1}), even if the delayed process itself involves nonlinear forces, $F$. Furthermore, their dynamical equations are deterministic, as opposed to $X_0$ which is subject to the white noise term $\xi$ [see~(\ref{EQ:Aux-Eq0})]. We stress that the set of auxiliary variables~(\ref{EQ:Interpretation_Xj}) is different from the one~(\ref{EQ:2timevarapp}) used in the earlier approach\red{, which we have reviewed in the preceding section}. A direct comparison is provided in Fig.~\ref{FIG:1}.%
\begin{figure}
\includegraphics[width=0.25\textwidth]{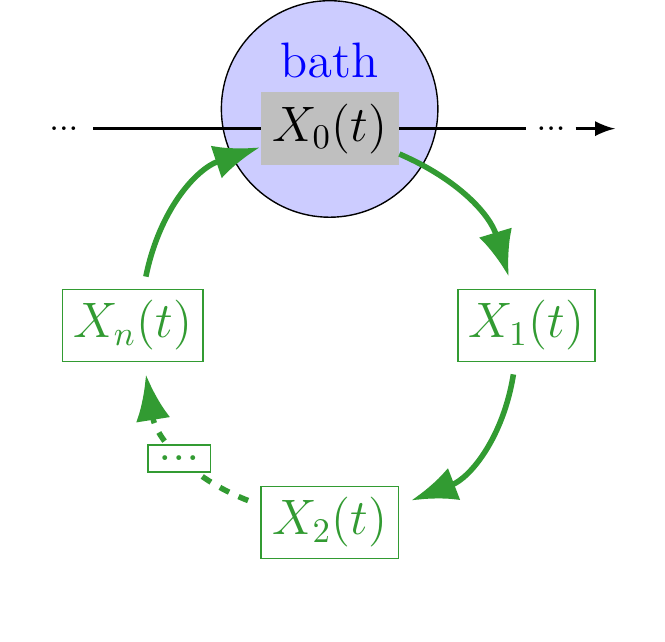}
\centering
\caption{Schematic visualization of the $(n+1)$-dimensional Markovian system described by Eqs.~(\ref{EQ:extended-LE}), with $n$ auxiliary variables. Only the variable $X_0$ is subject to noise, i.\,e., (thermal) fluctuations induced by a ``bath''.} %
\label{FIG:2}
\end{figure}

Next, we will use the Markovian embedding to find a probabilistic description of the delayed system.
\subsection{Fokker-Planck equation of the Markovian $(n+1)$-dimensional system} 
The Markovian $(n+1)$-dimensional system given by the set of coupled LEs~(\ref{EQ:extended-LE}), can also be described by the FPE
\begin{align}\label{EQ:FPE-extended-2}
\underbrace{ \partial_{t} \rho_{n +1} \white{\Bigg|}}_{\mathrm{(I)}} = \underbrace{- \partial_{x_0} \big[   F_{}(x_0,x_n) \rho_{n +1} \big] \white{\Bigg|}}_{\mathrm{(II)}}
\underbrace{- \frac{ n}{\tau}\sum_{j=1}^{n} \partial_{x_{j}}  \left[  (x_{j-1}-x_j) \rho_{n +1} \right] }_{\mathrm{(IV)}}
\underbrace{+ D_0 \, \partial_{x_{0}}^2  \rho_{n +1} \white{\Bigg|}}_{\mathrm{(III)}}
\end{align}
for the $(n+1)$-point (joint) PDF $\rho_{n +1}(x_0,t; x_1,t ; ... ;x_n,t )=\langle \delta[x_0-X_0(t)]\delta[x_1-X_1(t)]..\delta[x_n-X_n(t)] \rangle$. \red{Equation~(\ref{EQ:FPE-extended-2}) is a self-sufficient FPE, which describes the dynamics of the delayed process in a closed form in the limit $n\to\infty$.} 
\par
Importantly, $n$ will always be considered a \textit{finite} value here (until after marginalization). Finding a mathematical framework in which the limit $n\to \infty$ of Eq.\,(\ref{EQ:FPE-extended-2}) is meaningful and well-defined, is subject of ongoing work. Indeed, it is not clear how to formulate a Fokker-Planck equation on an infinite-dimensional space~\cite{Niculescu2012}. Moreover, one drift term is in fact proportional to $n$ itself, \red{making} this limit highly non-trivial. Here, we can safely ignore this problem, since we focus on the equations for the marginalized (low-dimensional) PDFs. As shown in Secs. \ref{SEC:FirstMember} and  \ref{SEC:HigherMember}, they can be derived by keeping $n$ finite and performing the limit only after the marginalization.

Equation~(\ref{EQ:FPE-extended-2}) resembles the FPE of a many-particle system. However, in sharp contrast to a system of interacting (colloidal) particles, the ``interactions'' are here \textit{unidirectional} (see Fig.~\ref{FIG:2}), which means that the ``particle'' $x_j$ only feels the ``particle'' $x_{j-1}$ and \textit{not }vice-versa. Hence, the forces between them are non-Newtonian (violating Newtons 3$^\mathrm{rd}$ law) and cannot be described by an interaction potential.
\section{Derivation of first member of Fokker-Planck hierarchy via Markovian embedding}\label{SEC:FirstMember}
\subsection{Marginalization}
In the following, we show that the delayed FPE for $\rho_1$ can be derived by a ``coarse-graining'' of~(\ref{EQ:FPE-extended-2}). To be more specific, we \textit{marginalize} the FPE~(\ref{EQ:FPE-extended-2}) for the joint PDF, that is we integrate it over the whole domain of all $n$ auxiliary variables, i.\,e., $\iiint_{\Omega} \mathrm{d}x_1\mathrm{d}x_2 .. \mathrm{d}x_n$, and, therewith obtain an equation for the marginal PDFs\footnote{We use the general relation between \textit{joint}, $\rho_2$, and \textit{marginal} distributions $\rho_{1}(z)=\int_{a} \rho_{2}(y,z) \mathrm{d}y$, with the normalization 
$\int_{b}\rho_{1}(z)\mathrm{d}z=\int_{b}\int_{a} \rho_{2}(y,z) \mathrm{d}y\mathrm{d}z=1$, with $y\in a$, $z\in b$.
}.
Importantly, we take the limit $n\rightarrow \infty$ only after the marginalization. We further use \textit{natural boundary conditions} at the endpoints of the domain $\Omega=[\omega_1,\omega_2]$ specified by
$$\lim_{x_j\to \omega_{1,2}} \rho_m(x_j,...)=\lim_{x_j\to \omega_{1,2}} \partial_{x_j}\rho_m(x_j,...)=0, ~~\forall j,m ,$$
which are justified in most physical scenarios. For sake of a shorter notation, we will omit the time arguments of the PDFs in lengthy expressions. The terms (I -- III) in Eq.~(\ref{EQ:FPE-extended-2}) can be readily marginalized to
\begin{subequations}
\begin{align}
\mathrm{(I)}&\rightarrow \partial_{t} \rho_1( x_0,t), ~~~~~~~
\mathrm{(II)}\to -  \partial_{x_0} \int_{\Omega}^{ }\left[ F_{}(x_0,x_n) \rho_{2}(x_0,t;x_n,t)\right]\mathrm{d}x_n, ~~~~~~~
\mathrm{(III)}\to \partial_{x_0}^2 \rho_1( x_0,t)
. \label{EQ:dFPE_rhs_1} 
\end{align}
The remaining drift terms $\mathrm{(IV)}$ in Eq.~(\ref{EQ:FPE-extended-2}) can be simplified by making use of their \textit{linearity} w.\,r.\,t. the various $x_j$ (please note that they are always linear as a result of the linearity in (\ref{EQ:Aux-Eq1}) despite of the fact that $F_{}$ is, in general, nonlinear). By application of the 
natural boundary conditions, and treating the remaining integrals by partial integration, we find
\begin{align}
\mathrm{(IV)}
\to& 
~\frac{ n}{\tau} \Bigg\{n \,\rho_{1}(x_0) -
\sum_{j=2}^{n} \iint_{\Omega}^{ }  x_{j} \, \partial_{x_j} \rho_{3}(x_0;x_{j-1};x_{j})  \mathrm{d}x_j \mathrm{d}x_{j-1}
+ \sum_{j=2}^{n} \int_{\Omega}^{ } x_{j-1} \underbrace{\left[\rho_{3}(x_0;x_{j-1}; x_{j})\right]_{\omega_1}^{\omega_2}}_{=0}  \mathrm{d}x_{j-1}
\nonumber \\
& 
 ~-\int_{\Omega}^{ } x_1 \, \partial_{x_1} \rho_{2}(x_0; x_1)  \mathrm{d}x_1
 +  x_{0} \underbrace{\left[\rho_{2}(x_0; x_1)\right]_{\omega_1}^{\omega_2}}_{=0}   
 \Bigg\}
\nonumber \\
&=\frac{ n}{\tau} \Bigg\{  n \,\rho_{1}(x_0)
+\sum_{j=2}^{n}\underbrace{\left[ x_{j} \rho_{3}(x_0;x_{j-1};x_{j})\right]_{\omega_1}^{\omega_2}}_{=0}  
-\sum_{j=2}^{n}\int_{\Omega}^{ } \rho_{2}(x_0;x_{j}) \mathrm{d}x_j 
-\underbrace{\left[ x_1 \rho_{2}(x_0; x_1)\right]_{\omega_1}^{\omega_2}}_{=0}  
-\rho_{1}(x_0) \Bigg\}
= 0.
 \label{EQ:dFPE_rhs_3}
\end{align}
Hence, the contribution to the drift term of the entire sum vanishes, and we obtain in total from (\ref{EQ:dFPE_rhs_1},\ref{EQ:dFPE_rhs_3}) 
\end{subequations}
\begin{align}\label{EQ:dFPE_1-member}
\partial_t \rho_1( x_0,t) = - \partial_{x_0} \int_{\Omega}^{ } F_{}(x_0,x_n) \rho_{2}(x_0;x_n)\mathrm{d}x_n + D_0 \partial_{x_0}^2 \rho_1( x_0,t).
\end{align}
So far, the marginalized FPE~(\ref{EQ:dFPE_1-member}) still pertains to a finite (but arbitrary) value of $n\in \mathbb{N}$. As a last step, we take {the limit}. To this end, we recall the meaning of the respective stochastic variables in this limit [Eq.~(\ref{EQ:Meaning_Aux})]. 
Based on that we perform the phase space transformation $\{x_0,t\} \rightarrow \{ x,t \} $ and $\{x_n,t\} \rightarrow \{ x_\tau,t-\tau \}$.
Therewith, we finally obtain the {delayed FPE} for the one-time PDF~(\ref{EQ:dFPE_start}). 

Next, we demonstrate that our procedure can also be applied to find higher members of the FP hierarchy.
%
\section{Derivation of higher members via Markovian embedding}\label{SEC:HigherMember}
{In the following, we derive a FPE for the two-time PDF $\rho_2(x,t;x_\tau,t-\tau)$. This is a particularly important quantity in the description of delayed processes, as it explicitly describes the correlation between the delayed and present state. This time, we marginalize the complete Fokker-Planck equation w.\,r.\,t. all auxiliary variables but the last one (again keeping $n$ finite until after the marginalization). We do not integrate over $x_n$ as this variable represents $X(t-\tau)$ in the limit $n\to \infty$. In a second step, we again perform a coordinate transformation to take the limit.}%
\subsection{Marginalization}
{
We again start from the FPE~(\ref{EQ:FPE-extended-2}) of the $n$-variable system and integrate the equation over $x_1,..,x_n$. 
Using similar steps as presented in Sec.\,\ref{SEC:FirstMember}, e.\,g., partial integration and subsequent application of the boundary conditions, we obtain the marginalized FPE}
\begin{align}
\label{EQ:dFPE_2nd-member_1}
 \partial_{t} \rho_2( x_0,t;x_n,t) =& -   \partial_{x_0}  \left[ F_{}(x_0,x_n) \rho_{2}(x_0,t;x_n,t)\right]
+
D_0 \partial_{x_0}^2\rho_2(x_0,t;x_n,t)
\nonumber \\&
 +  \partial_{x_n} \int_{\Omega }^{ } \frac{ n}{\tau}(x_{n}-x_{n-1}) \rho_{3}(x_0,t;x_{n-1},t;x_n,t) \mathrm{d}x_{n-1}
,
\end{align} %
which involves a three-point PDF. Please note that integrating (\ref{EQ:dFPE_2nd-member_1}) over $x_{n}\in\Omega$ yields the first member of the hierarchy~(\ref{EQ:dFPE_1-member}), which shows the consistency of our calculation.
Higher-order equations can be obtained in an analogous manner, yielding the infinite hierarchy of coupled equations
 \begin{align}
 \label{EQ:dFPE_3nd-member_1}
 \partial_t \rho_m(x_0 ;x_{n-m+2} ,..,x_n )= &-   \partial_{x_0}  \left[ F_{}(x_0,x_n) \rho_{m}(x_0;x_{n-m+2},..,x_n)\right]
 \nonumber \\&
  +\sum_{l=0}^{m-3}\frac{ n}{\tau}   \partial_{x_{n-l}}\left[(x_{n-l}-x_{n-l-1}) \rho_{m}(x_0;x_{n-m+2},..,x_n) \right]
 \nonumber \\&
  +\frac{ n}{\tau} \partial_{x_{n-m+2}} \int_{\Omega}^{ } (x_{n-m+2}-x_{n-m+1}) \rho_{m+1}( x_0;x_{n-m+1},..,x_n) \mathrm{d}x_{n-m+1} 
  \nonumber \\
   &
  +
  D_0\,  \partial_{x_0}^2\rho_m(x_0,t;x_{n-m+2},t,..,x_n,t), %
 \end{align} 
for $m\geq 3$. %
Note that in contrast to the FPE~(\ref{EQ:dFPE_1-member}) for $\rho_1$, all higher members of the FP hierarchy (\ref{EQ:dFPE_2nd-member_1},\ref{EQ:dFPE_3nd-member_1}) explicitly contain the linear forces stemming from the auxiliary variable Eqs.~(\ref{EQ:Aux-Eq1}).
\\
We now inspect in more detail the second member~(\ref{EQ:dFPE_2nd-member_1}) in the limit $n\to \infty$, to recover the system with discrete delay. We again need to transform the probability space, hence, the phase space and the PDFs defined on it.
\subsection{Transformation of phase space, limiting procedure}
In order to take the limit $n\to \infty$ of Eq.~(\ref{EQ:dFPE_2nd-member_1}), we will transform the phase space and the PDFs defined on it. %
We perform the transformation in two steps. First we replace the phase space variable $x_{n-1}$ by $\widetilde{x}= (n/{\tau})(x_{n-1} - x_{n})$, which will later in the limit $n\to\infty$ receive the meaning of a \textit{velocity}. In a second step, we transform back to the original variables of the delayed LE~(\ref{EQ:LE-orignial}), whereby mainly the time arguments will be affected. %

As first step, we change the phase space from $\{x_0,t;x_{n-1},t;x_n,t\}$ to $\{x_0,t;\widetilde{x},t;x_n,t\}$. To this end, we start with performing the {substitution} $x_{n-1} ~\longrightarrow ~\widetilde{x}= (n/{\tau})(x_{n-1} - x_{n})$ to rewrite the integral $\int_{\Omega}^{ } \mathrm{d}x_{n-1} $ as $({\tau}/{n})\int_{ \widetilde{\Omega}}^{ } \mathrm{d}\widetilde{x}$.
\begin{subequations}
The drift term then becomes
\begin{align}
 \frac{ n}{\tau}\partial_{x_n} \int_{\Omega}^{ } (x_{n}-x_{n-1})  &\rho_{3}(x_0,t;x_{n-1},t;x_n,t) \mathrm{d}x_{n-1}
= \frac{ \tau}{n}  \partial_{x_n}\int_{\widetilde{\Omega}}^{ } \widetilde{x} \,\rho_3\left(x_0,t;\frac{\tau}{n}\widetilde{x}+ x_{n},t; x_{n},t\right) \mathrm{d}\widetilde{x}.
\end{align}
In the next step, 
we need to transform the PDF $\rho_3\to \widetilde{\rho_3}$, where $\widetilde{\rho_3}$ is a normalized (joint) PDF of the new phase space variables $\{x_0,\widetilde{x},x_n\}$.
The relation between both PDFs can be found by comparing the normalizations. In particular,
\begin{align}
\iiint_{\Omega}^{ }{\rho_3}\left(x_0,t; x_{n-1},t; x_{n},t\right)\mathrm{d}{x_0}\mathrm{d}{x}_{n-1}\mathrm{d}{x_n} 
=\frac{\tau}{n}\int_{\Omega}^{}\int_{\widetilde{\Omega}}^{ }\int_{\Omega}^{ } {\rho_3}\left(x_0,t;\widetilde{x}+ x_{n},t; x_{n},t\right)\mathrm{d}{x_0}\mathrm{d}\widetilde{x}\mathrm{d}{x_n} &= 1,
\end{align}
and, on the other hand, by definition,
$\iiint_{}^{ }\widetilde{\rho_3}\left(x_0,t;\widetilde{x},t; x_{n},t\right)\mathrm{d}{x_0}\mathrm{d}\widetilde{x}\mathrm{d}{x_n} = 1$.
Combining both conditions yields
\begin{align}{\rho_3}\left(x_0,t; \frac{\tau}{n}\widetilde{x}+ x_{n},t; x_{n},t\right)\stackrel{!}{=}\frac{\tau}{n}\widetilde{\rho_3}\left(x_0,t;\widetilde{x},t; x_{n},t\right).\end{align}
Therewith, we obtain the drift term
\begin{align}
\frac{ n}{\tau} \partial_{x_n} \int_{\Omega}^{ } (x_{n}-x_{n-1})  \rho_{3}(x_0;x_{n-1};x_n) \mathrm{d}x_{n-1}
\longrightarrow&  \partial_{x_n}\int_{\widetilde{\Omega}}^{ } \widetilde{x} \,\widetilde{\rho_3}\left(x_0;\widetilde{x}; x_{n}\right) \mathrm{d}\widetilde{x}
\\
=&\partial_{x_n} \left[ \left\langle \widetilde{X} ( t) \big|x_0=X_0(t) , x_n=X_n(t)\right \rangle\rho_2(x_0,t;x_n,t) \right]
.
\end{align}
\\
In the last line, we have used the conditional average notation, which involves the respective stochastic variables. Now, we recall the meaning of the stochastic variables in the limit $n\rightarrow \infty$, i.\,e., $X_0(t) = X(t)$, $X_n(t)= X(t-\tau) $ from~(\ref{EQ:Meaning_Aux}), and $\dot{X}_n(t)=\frac{n}{\tau}\left[ X_{n-1}(t) - X_{n}(t) \right]$ from~(\ref{EQ:Aux-Eq1}).
Based on this, we perform the transformation
\begin{align}
\{x_0,t\} &\rightarrow \{ x,t \} ,\\
\{x_n,t\} &\rightarrow \{ x_\tau,t-\tau \} ,\\
\{\widetilde{x},t\} &\rightarrow \{ \dot{x}_\tau,t-\tau \},\\
\partial_{x_n}\int_{\widetilde{\Omega}}^{ } \widetilde{x} \,\widetilde{\rho_3}\left(x_0,t;\widetilde{x},t; x_{n},t\right) \mathrm{d}\widetilde{x} &\rightarrow 
\partial_{x_\tau}\left[ \left\langle \dot{X} ( t - \tau ) \big|x=X(t) , x_\tau=X({t - \tau})\right \rangle\rho_2(x,t;x_\tau,t-\tau) \right], \label{EQ:Drift-term-No-Vel}
\end{align}
\end{subequations}
with $ \dot{x}_\tau\in 
\widetilde{\Omega}=\mathbb{R}$. 
We note that the appearance of a velocity-like variable is unusual, as we deal with overdamped dynamics. 
Since $\dot{X}$ has no well-defined corresonding PDF for overdamped dynamics~\cite{Gardiner2002}, we omit the notation used on the left side of~(\ref{EQ:Drift-term-No-Vel}).
With these steps, we finally obtain from~(\ref{EQ:dFPE_2nd-member_1}), which is the FPE for the two-point PDF of the Markovian system~(\ref{EQ:extended-LE}), the {second member} of the {FP hierarchy} for the delayed system
\begin{align}
\partial_t \rho_2(x,t;x_\tau,t-\tau) =& - { \partial_x}  \left[ F_{}(x,x_\tau) \rho_{2}(x,t;x_\tau,t-\tau)\right]
 +
D_0  \partial_{x}^2 \rho_2(x,t;x_\tau,t-\tau)
\nonumber \\&
 %
 + \partial_{x_\tau} \left[\langle \dot{X} ( t - \tau ) |x=X(t) , x_\tau=X({t - \tau})\rangle \rho_2(x,t;x_\tau,t-\tau)  \right]
%
\label{EQ:dFPE_2nd-member_2}
\end{align}
Comparison with the corresponding FPE~(\ref{EQ:General_dFPE2}) obtained from Novikov's theorem reveals that Eq.~(\ref{EQ:dFPE_2nd-member_2}) is in fact quite different. Our approach thus yields an \textit{alternative representation} of the second member. However, we show in the Appendix~\ref{APP:Nov-Embedd-Equivalence}, that one can indeed transform~(\ref{EQ:dFPE_2nd-member_2}) into (\ref{EQ:General_dFPE2}). \red{As we will argue below,} this alternative representation is a new starting point to approximate the two-time PDF, $\rho_2(x,t;x_\tau,t-\tau)$, \red{for which, to our knowledge, no approaches have been reported}. Moreover, also the Eqs.~(\ref{EQ:dFPE_2nd-member_1},\ref{EQ:dFPE_3nd-member_1}) might potentially serve for this purpose.

{The main difference of~(\ref{EQ:dFPE_2nd-member_2}) as compared to~(\ref{EQ:General_dFPE2}) (from Novikov's theorem), is that it involves 
cross-correlations of the present and delayed system state with the delayed \textit{velocity}, instead of the correlations among the system states at three different times [via the
three-{time} PDF $\rho_3(x,t;x_\tau,t-\tau,{x_{2\tau}, t-2\tau})$]. This indicates that, instead of going back even further \textit{into the past} with every member [by first taking into account $X(t-\tau)$, then $X(t-2\tau)$, then $X(t-3\tau)$ ...], }the here derived hierarchy rather collects more and more information about the dynamics at time $t-\tau$ [by taking into account $X(t-\tau)$, then $\dot{X}(t-\tau)$]. 

Both representations are valid and describe the same process.
This shows that to (probabilistically) predict the future after a time $t$, either knowledge of the system states at all times $t,t-\tau,t-2\tau,...$ would be needed, or, complete knowledge about the preceding interval $[t-\tau,t]$ would suffice. On the stochastic level of description, the process which underlies the non-Markovian one, can either be considered as a process of the state vector $(X(t),X(t-\tau),X(t-2\tau),X(t-3\tau),..)$, or, equivalently (and actually more natural), of the state vector of all $X(s)$ $\in \mathbb{R}^\infty$, with $s\in [t-\tau,t)$. \blue{Moreover, the here presented results show that the ``discretization of the past'' (i.\,e., the discretization of $[t-\tau,t)$ into $n\in \mathbb{N}$ slices), as implied by the employed embedding technique, is a valid approach to time-continuous dynamics with delay.}
\par
A technical advantage of the representation~(\ref{EQ:dFPE_2nd-member_2}) is that it does not involve unknown functional derivatives w.\,r.\,t. the noise, as opposed to~(\ref{EQ:General_dFPE2}), and only contains one diffusion term w.\,r.\,t. $x$ (and no additional diffusion term in $x_\tau$). 
\red{In the next section, we will discuss an approximation scheme based on~(\ref{EQ:dFPE_2nd-member_2}).}
\section{\red{Approximation for the two-time PDF}}\label{SEC:Approx}
\red
{In this section, we discuss the possibility of finding an approximate two-time PDF based on~(\ref{EQ:dFPE_2nd-member_2}). 
We are particularly interested in an approach that is justified when the delay time $\tau$ is comparable with other timescales of interest, because in this regime, the non-Markovian effects are most prominent. We further focus on non-equilibrium steady states, which are typically approached \blue{after the transient behavior due to the initial condition has decayed}. {By making an ansatz for the higher order correlations,} we find from~(\ref{EQ:dFPE_2nd-member_2}) a closed equation that can be solved analytically.}
\par
\red{As a first step, we assume that, \blue{if the delay time is not very small}, the correlations between the system state $X(t)$ and the delayed displacement $\mathrm{d}X(t-\tau)$ can be neglected, as well as the correlations between system state and instantaneous displacement. This amounts to the ansatz}
\red{
$ \langle \dot{X} ( t - \tau ) |X(t), X({t - \tau} )\rangle \approx \langle \dot{X} ( t - \tau ) \rangle $.
Inserting this into~(\ref{EQ:dFPE_2nd-member_2}), one obtains}
\begin{align}
- \langle \dot{X} \rangle\,\partial_{x_\tau}\rho_2(x,t;x_\tau,t-\tau)=& - { \partial_x}  \left[ F_{}(x,x_\tau) \rho_{2}(x,t;x_\tau,t-\tau)\right]
 +
D_0  \partial_{x}^2 \rho_2(x,t;x_\tau,t-\tau),
\label{EQ:dFPE_2nd-member_2-MF}
\end{align}
involving the particle current $\langle \dot{X} \rangle$. In order to find an ansatz for the latter, one can proceed by utilizing the framework~\cite{Reimann2002} (developed for Markovian systems), which gives a connection between $\langle \dot{X} \rangle$ and the probability current $J$ via 
$\langle \dot{X} \rangle= \int_{\Omega} J(x,t) \mathrm{d}x$, where $J$ is found from the FPE $\partial_t \rho_1 = \partial_x J$ (written in the form of a continuity equation). For the delayed systems considered here $J(x,t)= \int_{\Omega} F(x,x_\tau) \rho_2(x,x_\tau) \mathrm{d}x_\tau -D_0 \partial_x \rho_1(x)$. These assumptions give rise to the approximate, closed FPE~(\ref{EQ:dFPE_2nd-member_2-MF}) with $\langle \dot{X} \rangle= \iint_{\Omega} F(x,x_\tau) \rho_2(x,x_\tau) \mathrm{d}x_\tau\mathrm{d}x$\blue{. We expect the approximation to break down when $\tau$ is very small, i.\,e., when the correlations between $X(t)$ and $\mathrm{d}X(t - \tau)$ are not negligible.}
\par
\red{For a common type of systems, this approximate equation becomes even simpler without requiring further assumptions or simplifications, that is, when $\langle \dot{X} \rangle =0$ in the steady-state. For example, this is always the case for systems that obey a symmetry w.\,r.\,t. coordinate inversions $x \to -x$. Then,~(\ref{EQ:dFPE_2nd-member_2-MF}) has the formal solution
\begin{equation}
\rho_2(x,t;x_\tau,t-\tau) = Z^{-1} h(x_\tau) e^{\int F(x,x_\tau) \mathrm{d}x},
\end{equation}
where $Z$ is a normalization constant. Further, $h(x_\tau)$ is a function that is not determined by~(\ref{EQ:dFPE_2nd-member_2-MF}) alone, but is, however, determined by the steady-state constraint $\partial_t \rho_1 =0$, which implies $ \rho_1(x)=\rho_1(x_\tau)\Rightarrow \int_\Omega \rho_2(x,x_\tau) \mathrm{d}x = \int_\Omega \rho_2(x,x_\tau) \mathrm{d}x_\tau$.}
\par
\red{As an important example, we consider systems with \textit{linear }delay force, and (inversion-symmetric) nonlinear, nondelayed force $F_\mathrm{nd}$, i.\,e., $F(x,x_\tau)=F_\mathrm{nd}(x)+k x_\tau$. Then 
$h= e^{\int F_\mathrm{nd}(x_\tau) \mathrm{d}x_\tau}$ yields
\begin{equation}\label{EQ:Approx_general}
\rho^{\mathrm{approx}}_2(x,t;x_\tau,t-\tau) = Z^{-1} e^{\int F_\mathrm{nd}(x) \mathrm{d}x+kx x_\tau+\int F_\mathrm{nd}(x_\tau) \mathrm{d}x_\tau},
\end{equation}
which fulfills~(\ref{EQ:dFPE_2nd-member_2-MF}) and the steady-state constraint.}

\red{A peculiar feature of~(\ref{EQ:Approx_general}) is that it does not explicitly depend on the precise value of $\tau$ (unless $F_\mathrm{nd}$ does). However, it has been observed in earlier studies for both, nonlinear~\cite{Loos2017} and linear systems~\cite{Kuechler1992} (where the PDFs can be computed analytically), that the moments and correlation functions show a saturation in $\tau$. In other words, $\rho_1$ and $\rho_2$ become in fact independent of $\tau$ in the regime of large delay times. Since we expect our approximation to be justified only in the large $\tau$ regime, we suspect that we can understand the independence of~(\ref{EQ:Approx_general}) on $\tau$ in this sense. We recall that the approximation is expected to break down for small values of $\tau$, because then the correlations between $\dot{X}(t-\tau)$ and $X(t)$ are non-negligible.}
\subsection{\red{Application to a bistable delayed system}}
\red{To test the approximation~(\ref{EQ:Approx_general}), we consider a concrete example, namely a 
bistable system~\cite{Tsimring2001,Loos2019,Loos2017} given by the rescaled LE\footnote{The dimensionless LE is obtained by rescaling position $x/\sigma \to x$ and time $(D_0/\sigma^2)t \to t$, $(D_0/\sigma^2)\tau \to \tau$, with the the minima $x\pm \sigma$ of the doublewell potential $V=V_0 [(x/\sigma)^4-2(x/\sigma)^2]$, and by rescaling the parameters $V_0/(\gamma D_0) \to V_0 $ and $k/(\gamma D_0) \to k$ with the bath's thermal energy $\gamma D_0$.}
\begin{equation}\label{EQ:LE-bistable}
\dot{X}(t) = -4V_0\left[X(t)^3-X(t)\right] + k \left[X(t-\tau)-X(t)\right] +\sqrt{2} \,\xi(t).
\end{equation}
Equation (\ref{EQ:LE-bistable}) describes the dynamics of a stochastic system in a static doublewell potential $V=V_0 (x^4- 2x^2)$ with minima at $x\pm 1$, subject to a linear delay force stemming from an ``optical tweezers'' potential $(k/2) (x- x_\tau)^2$~\cite{Kotar2010}.}
\red{Using Eq.~(\ref{EQ:Approx_general}), we can readily write down the approximate steady-state two-time PDF
\begin{align}\label{EQ:Approx_bistable}
\rho_2(x,t;x_\tau,t-\tau) = Z^{-1} e^{-V_0 (x^4- 2x^2) -\frac{k}{2}x^2 -V_0 (x_\tau^4- 2x_\tau^2) -\frac{k}{2}x_\tau^2  +kx x_\tau} . 
\end{align}
}
%
%
%
%
%
\begin{figure}
\includegraphics[width=.8\textwidth]{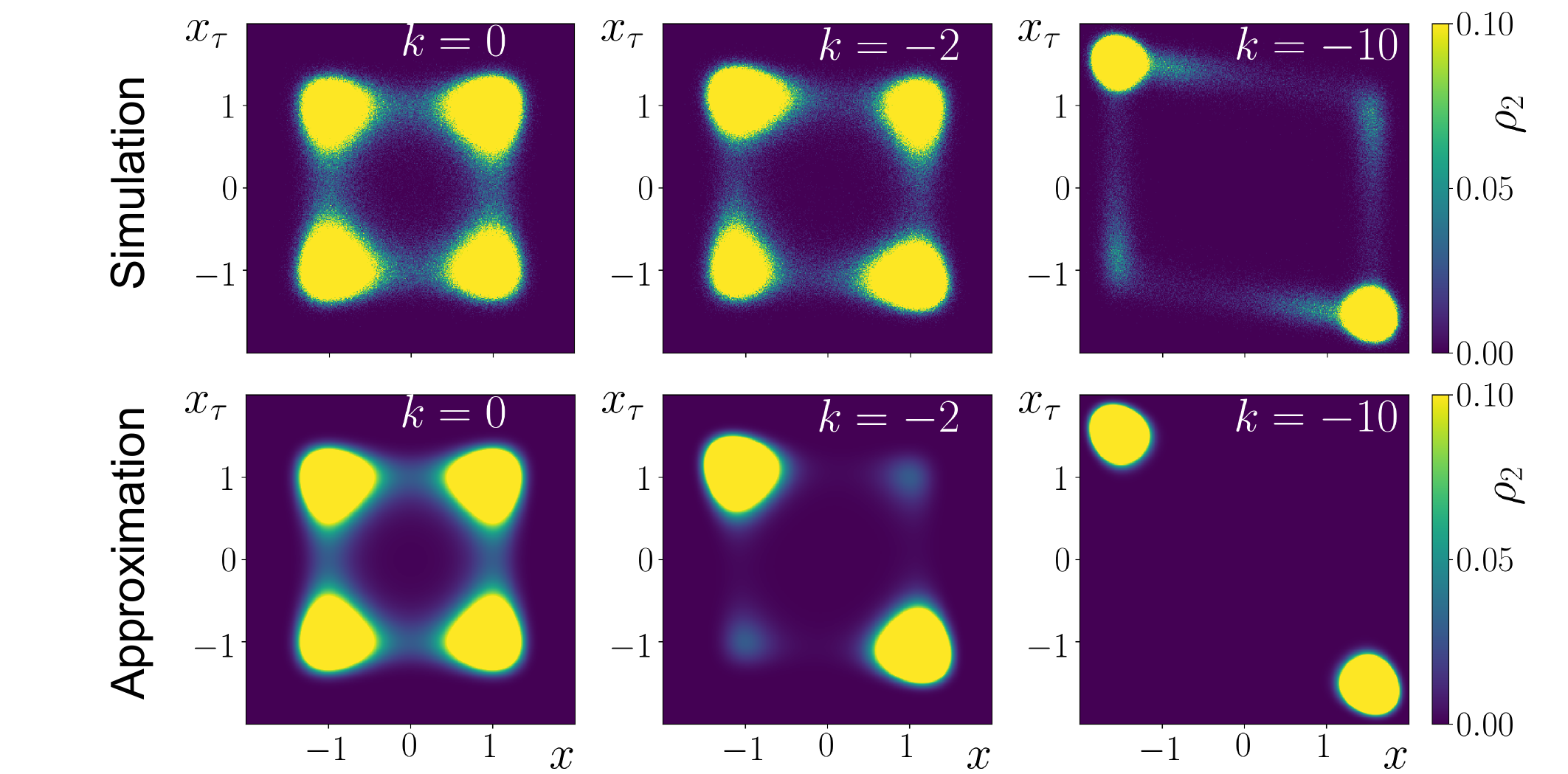}
\caption{Two-time PDF in the doublewell potential with linear delay force $k X(t-\tau)$,  with increasing magnitude of negative $k$, \textit{upper panels}: numerical results from LE~(\ref{EQ:LE-bistable}), \textit{lower panels}: approximated $\rho_2$ from Eq.~(\ref{EQ:Approx_bistable}). Other parameters $V_0=3.5$, $\tau=10$.}\label{FIG:3}
\end{figure}
\red{It is well-known that the bistable system~(\ref{EQ:LE-bistable}) exhibits delay-induced oscillations between the two potential minima, when the delay time is larger than the the typical time between noise-induced jumps, and when $|k|$ has the same order of magnitude as $V_0$. Because these jumps manifest in the two-time PDF, as will be explained below, we consider this parameter regime %
(we fix $V_0\!=\!3.5$ and $\tau\!=\!10$) and only vary $k$, particularly focusing on the question whether the approximation can capture the delay-induced jumps.
To test the results from a quantitative point of view, we have additionally performed direct Brownian dynamics simulations of Eq.~(\ref{EQ:LE-bistable}) with $>2000$ realizations, temporal discretization $10^{-4}$, and cutoff $>10^{6}$ time steps to reach steady-state conditions.
} 
\par
\red{
First, we consider the system in the absence of the delay force ($k=0$), where the dynamics is dominated by the noise, and the probability to find $X$ is always highest around the potential minima $x=\pm 1$. As thermal jumps are very likely on the timescale of $\tau$, the chance of finding the system in one well is independent of the question whether it was on the same or on the other side $\tau$ ago, meaning that the probabilities to find $X(t)$ and $X(t-\tau )$ are independent from each other. 
We hence expect the joint two-time PDF\footnote{Due to the symmetry of the external potentials, we further always expect that $\rho_2(x,x_\tau)$ is symmetric w.\,r.\,t.\, $\{x,x_\tau\} \to -\{x,x_\tau\} $.} $\rho_2(x,x_\tau)$ to have pronounced peaks of identical heights around $\{x,x_\tau \} \approx \{\pm 1,\pm 1 \}$ and $\{\pm 1,\mp 1 \}$. This expectation is confirmed by numerical results, and is correctly described by our approximation~(\ref{EQ:Approx_bistable}), as plotted in Fig.\,\ref{FIG:3} (left plots, upper and lower panel, respectively). (We note that for $k = 0$ the dynamics is Markovian and the system asymptotically approaches thermal equilibrium.) We now consider the impact of $k\neq 0$ (where a non-equilibrium steady-state is approached).
}
\par
\red{
For negative $k<0$, i.\,e, when the linear feedback force is directed towards the delayed system state, the induced oscillations are known to have a mean period of roughly $2\tau$~\cite{Loos2017,Loos2019,Tsimring2001}. This means that the jumps {reduce} the joint probability to find the system in the same well at times $t-\tau$ and $t$. We hence expect the delay-induced oscillations to lower the two-time PDF around $\{\pm 1,\pm 1 \}$, and increase it around $\{\pm 1,\mp 1 \}$. This effect is indeed seen in both, the numerically obtained PDF and the approximate results from~(\ref{EQ:Approx_bistable}), see Fig.\,\ref{FIG:3}.
Interestingly, by changing the sign of $k$, the mean period of the oscillations can be switched to about $\tau$~\cite{Tsimring2001}. Thus, we expect the delay-induced oscillations to have, in fact, the opposite effect for positive $k$, i.\,e., enhance the PDF at $\{\pm 1,\pm 1 \}$, and reduce it around $\{\pm 1,\mp 1 \}$. Again, this expected behavior is confirmed by simulation results, and as well captured nicely by the proposed approximation, see Fig.\,\ref{FIG:4}.
}
\par
\red{
The comparison between numerical results and the approximate $\rho_2$ from~(\ref{EQ:Approx_bistable}) in Figs.\,\ref{FIG:3}, \ref{FIG:4} includes a broad range of $k$ values.
Remarkably, we find that the proposed approximation yields very good results in all cases considered. In particular, it captures the main characteristics of the two-time PDF described above, for negative and for positive $k$. Even quantitatively, the agreement is very reasonable. We further considered higher barrier heights ($V_0=5$), smaller delays ($\tau=5$), as well as higher feedback strength (up to $|k|=20$) and found again very good agreement (not shown here). On the other hand, the approximation breaks down if the delay time is very short (we have tested, e.\,g., $\tau=0.1$). We suspect that in this regime, the correlations between $\dot{X}(t-\tau)$ and $X(t)$ are not negligible (hence, the assumption 
{$ \langle \dot{X} ( t - \tau ) |X(t),X({t - \tau})\rangle \approx \langle \dot{X} ( t - \tau ) \rangle $}
is not justified).
}
%
%
%
%
\begin{figure}
\includegraphics[width=.8\textwidth]{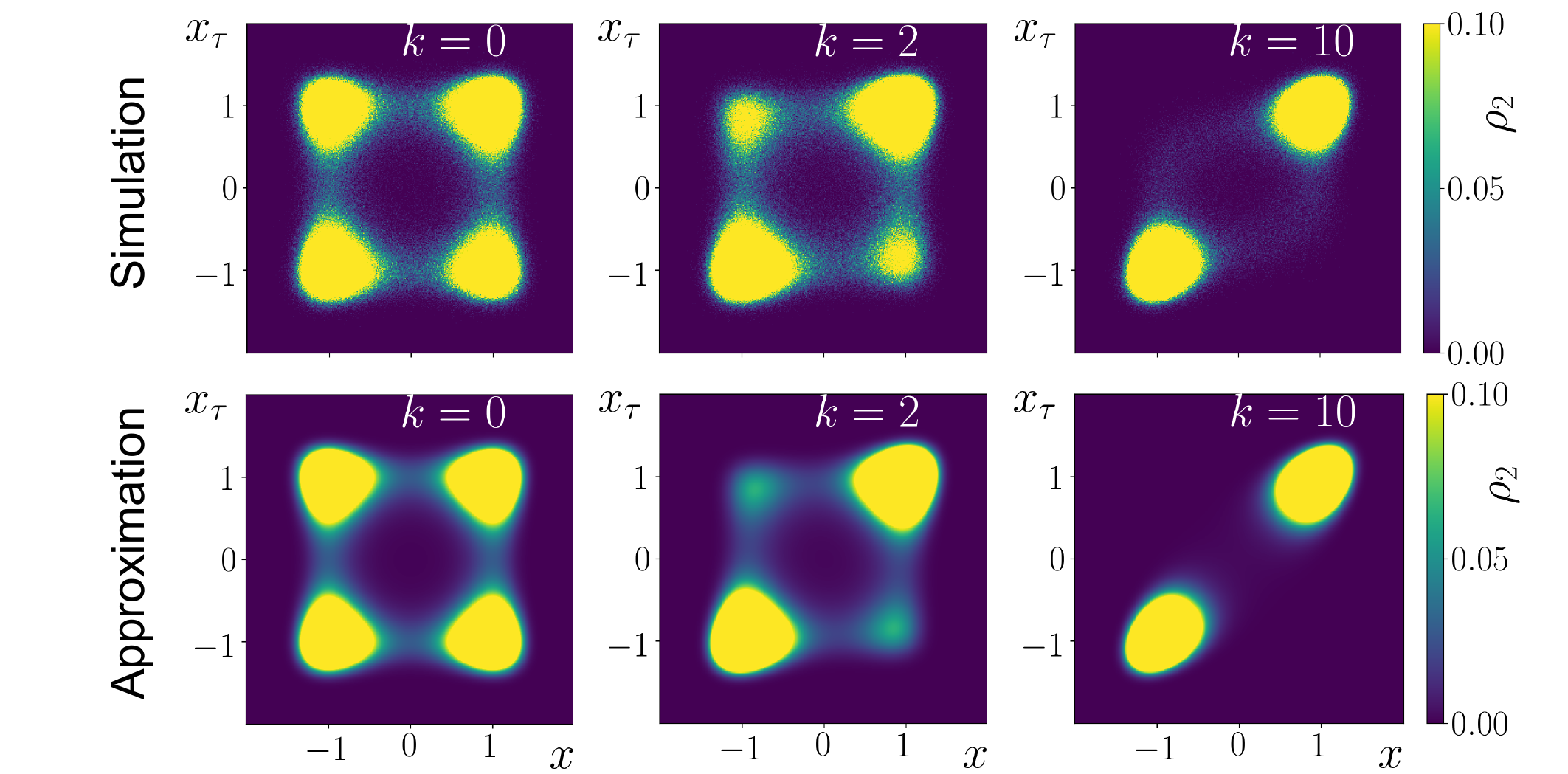}
\caption{Two-time PDF in the doublewell potential with linear delay force $k X(t-\tau)$, with increasing magnitude of {positive} $k$, \textit{upper panels}: numerical results from LE~(\ref{EQ:LE-bistable}), \textit{lower panels}: approximated $\rho_2$ from Eq.~(\ref{EQ:Approx_bistable}). Other parameters $V_0=3.5$, $\tau=10$.}
\label{FIG:4}
\end{figure}
%
%
%
%
\par
\red{
The accuracy of the approximate $\rho_2$ is indeed a remarkable result, in particular, because the underlying assumption appears to be rather crude. This demonstrates that an approximation scheme on the level of the second member of the FP hierarchy is somewhat superior as compared to an approach on the level of the first member. To further support this argument, we compare the prediction of approximation~(\ref{EQ:Approx_bistable}) to earlier approaches for the one-time PDF known from the literature (to the best of our knowledge, approaches for the two-time PDF have not been presented elsewhere). To this end, we marginalize $\rho_2$ from Eq.~(\ref{EQ:Approx_bistable}), i.\,e., $\rho_1(x)=\int_{\Omega} \rho_2(x,x_\tau)\mathrm{d}x_\tau$ and compare the result with the small $\tau$ expansion from~\cite{Guillouzic1999}, and the force-linearization closure (FLC)~\cite{Loos2017}. As can be seen in Fig.\,\ref{FIG:5},
the novel approximation yields the best results in the regime of large delay times. 
Admittedly, these other approaches are, in fact, known to be not justified for large $\tau$ values, or when the delay-induced jumps are very likely, respectively~\cite{Loos2017}. We nevertheless use them to compare with, due to the lack of other approximation schemes for this nonlinear system. \blue{In Appendix\,\ref{APP:Short-tau}, a short delay times is considered, where the other approximations are more reasonable.}
}
\par
\red{
We note that besides the features discussed above, the numerically obtained two-time PDFs in Figs.\,\ref{FIG:3} and \ref{FIG:4} show very interesting further details. An example is the broken symmetry $\rho_2(a,t; a',t-\tau) \neq \rho_2(a',t,a,t-\tau)$, which is not captured by the approximation~(\ref{EQ:Approx_bistable}). A study of these features is beyond the scope of this work, but it would be interesting to see how these details are connected with dynamical and thermodynamical properties of the non-equilibrium steady-state.
}
%
%
%
%
%
\begin{figure}[!htb]
\begin{minipage}{0.45\textwidth}
{\large (a) $k=-2$, $\tau=10$}\\
\includegraphics[width=.85\textwidth]{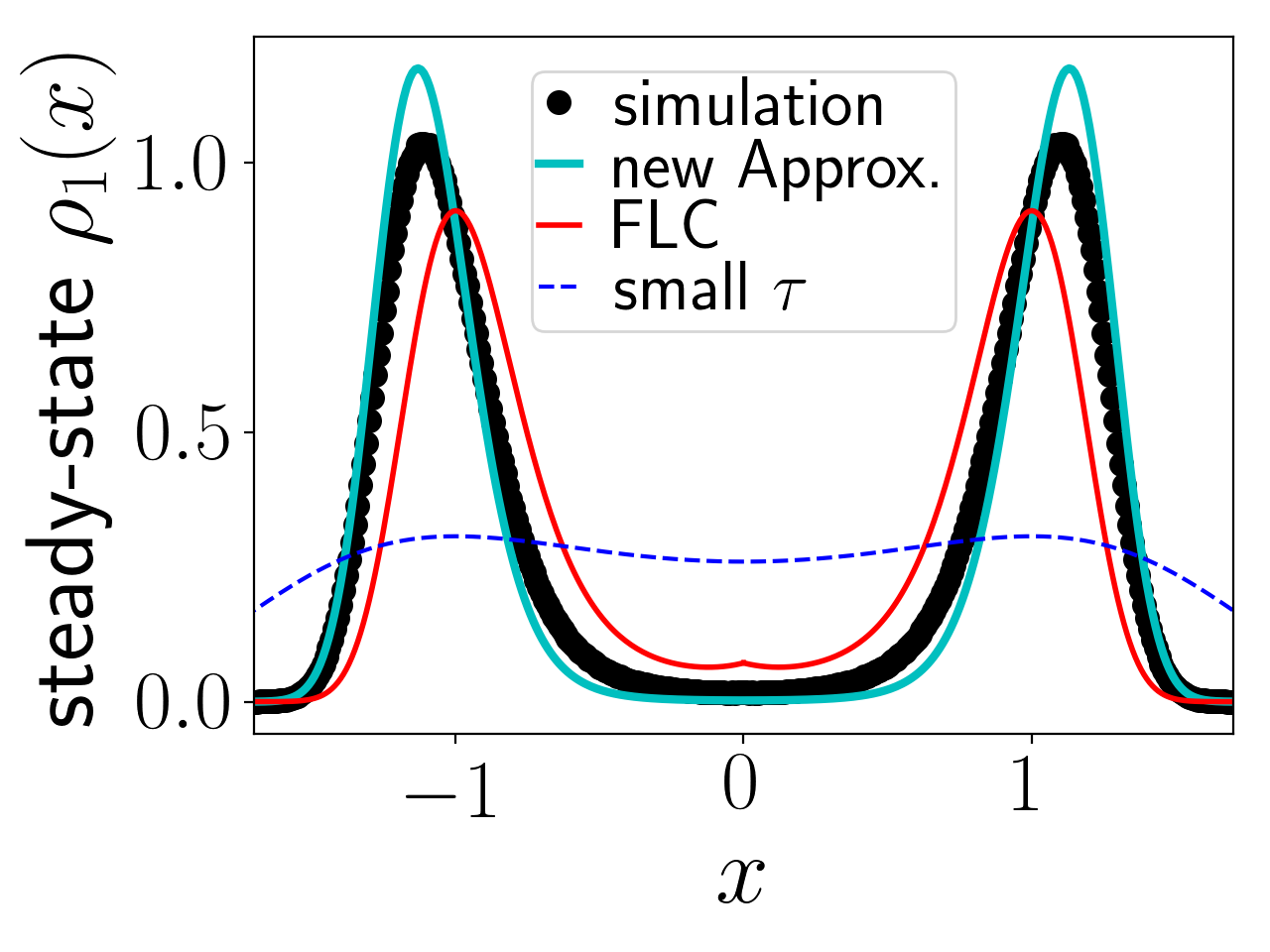}
\\
{\large (c) $k=2$, $\tau=10$}\\
\includegraphics[width=.85\textwidth]{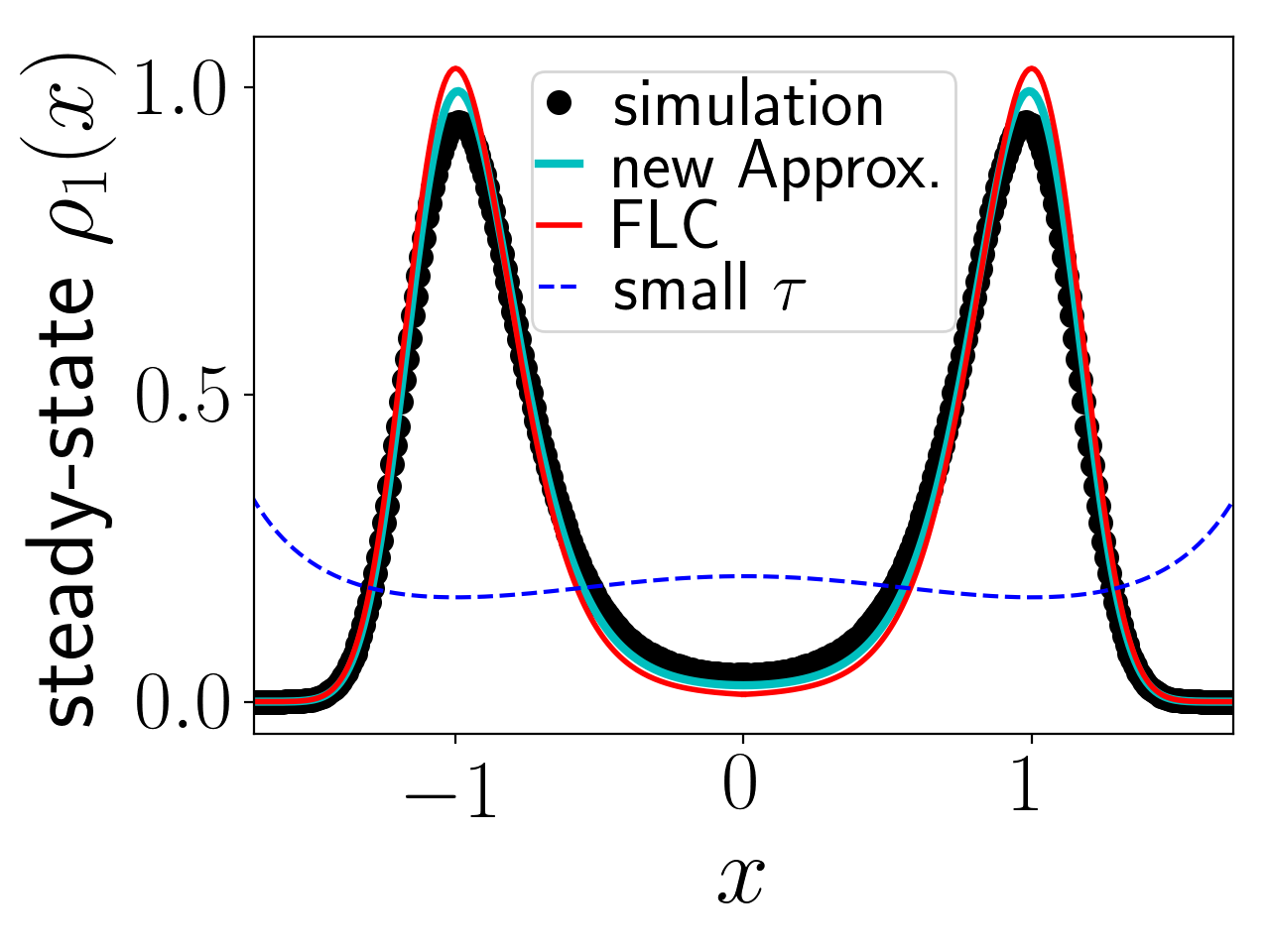}
\end{minipage}
\hfill
\begin{minipage}{0.45\textwidth}
{\large (b) $k=-10$, $\tau=10$}\\
\includegraphics[width=.85\textwidth]{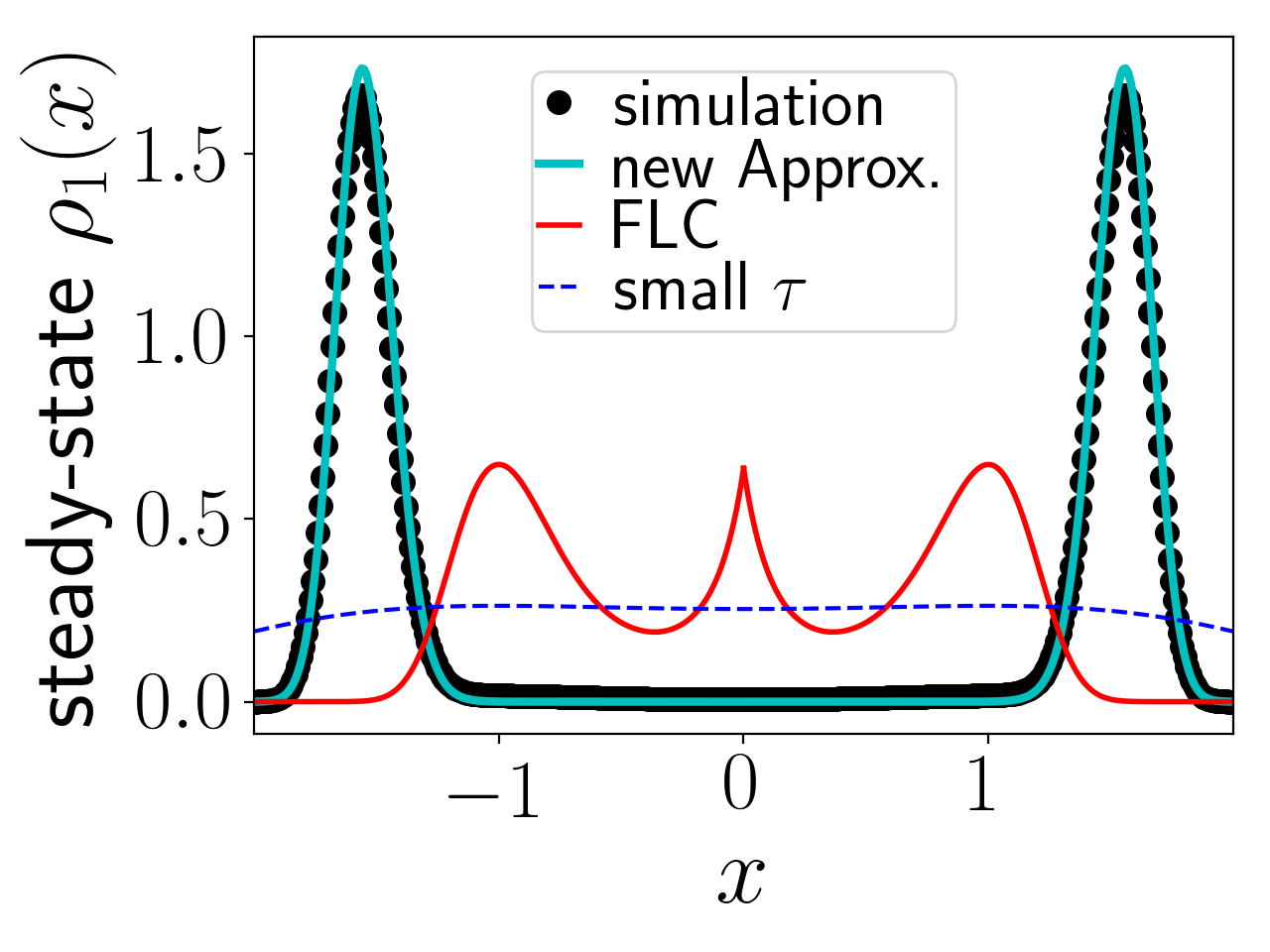}
\\
{\large (d) $k=10$, $\tau=10$}\\
\includegraphics[width=.85\textwidth]{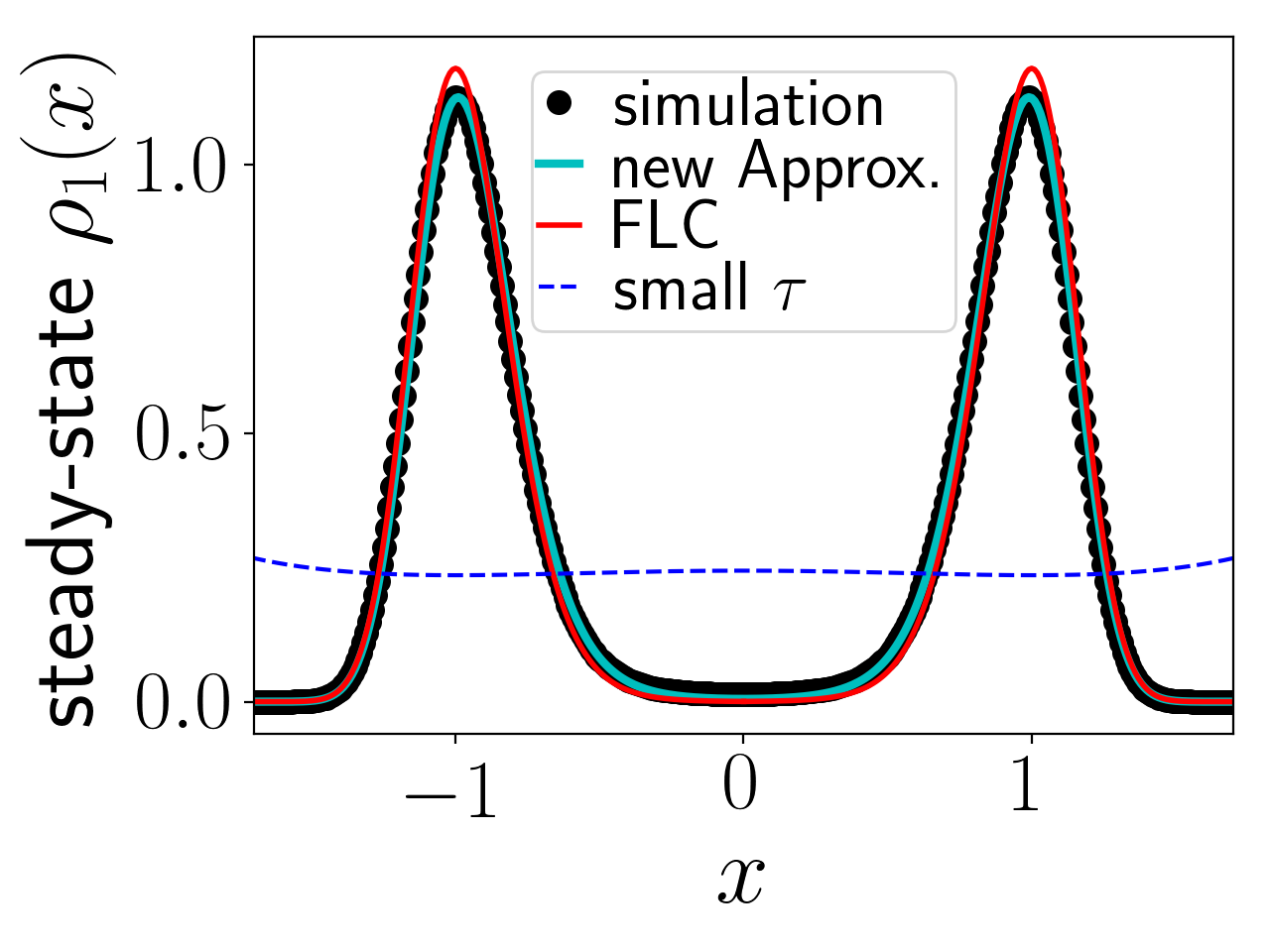}
\end{minipage}
\caption
{
Comparison of different approximations for the one-time PDF in the doublewell potential with linear delay force $k X(t-\tau)$, at $V_0=3.5$, $\tau=10$, and at 
various values of $k$;
\textit{blue dashed lines:} small $\tau$ approximation (which is only justified for very small $\tau$), \textit{cyan lines:} the new approximation $\int_{\Omega} \rho_2(x,x_\tau)\mathrm{d}x_\tau$ with $\rho_2$ from Eq.~(\ref{EQ:Approx_bistable}), \textit{red lines:} Force-linearization closure FLC from~\cite{Loos2017}. \textit{black disk}: Brownian dynamics simulations. \blue{The cases $k=2$, $\tau=0.1$, and $k=-10$, $\tau=0.1$ are considered in Appendix\,\ref{APP:Short-tau}.}
}\label{FIG:5}
\end{figure}
%
\section{Conclusions}
In this work, we have discussed the probabilistic description of delayed stochastic systems. As a starting point, we have reviewed an earlier approach based on Novikov's theorem, from which a Fokker-Planck description can be derived in the form of an infinite hierarchy~\cite{Frank2005a,Loos2017,Rosinberg2015}. The first member is the well-known FPE for the one-time PDF, which contains the two-time PDF and is thus not closed. Still, this equation has in the past proven itself to be an important tool in the search for exact results~\cite{Frank2001,Loos2019} and a valuable base for approximations~\cite{Loos2017,Frank2005}. The main purpose of our work was to shed light onto the higher members of this hierarchy, which have rarely been discussed in earlier literature.

To this end, we introduced an alternative FP description basing on a Markovian embedding technique, where the delayed process is represented as the dynamics of one variable of an $(n+1)$-variable Markovian system.
By projecting the corresponding closed, Markovian FPE onto lower-dimensional subspaces, and taking the limit $n\to \infty$ in the projected equations, we derived a hierarchy of FPEs. While the first member is identical, the representation of the higher members derived in this way differs from the one obtained by Novikov's theorem. \blue{We demonstrated that both hierarchies can, however, be converted into each other. This shows that a ``discretization of the past'' (i.\,e., the discretization of $[t-\tau,t)$ into $n\in \mathbb{N}$ equidistant slices), which is implied by the employed embedding technique, is a valid approach to time-continuous dynamics with delay.} Furthermore, the higher members derived here [Eqs.~(\ref{EQ:dFPE_2nd-member_1},\ref{EQ:dFPE_3nd-member_1}) or (\ref{EQ:dFPE_2nd-member_2})], might serve as a new starting point for approximations. 
\par
\red{For the second member, which we focused on in this paper, we have proposed one possible approximation scheme, and %
applied %
it to a bistable model with linear delay force. We found remarkably good agreement with simulation results, despite the simple form of the obtained approximate two-time PDF. With this, we present for the first time approximate results for this quantity which is indeed central in the probabilistic description of stochastic systems with time delay. %
The proposed scheme renders very good results in the regime of large delay times, for both cases, feedback forces that are directed towards, or away from the delayed position. In particular, it captures the main features of the two-time probability density which has very distinct characteristics depending on the direction of the feedback, related to delay-induced oscillations. Moreover, the corresponding approximate one-time PDF (which can be obtained by marginalization) outperforms earlier approaches known from the literature. %
}\par
The benefit of an approximation schemes based on Eqs.~(\ref{EQ:dFPE_2nd-member_2}) or~(\ref{EQ:General_dFPE2}), as compared to an approach operating on the first member (like the ones known from the literature), is twofold. Besides expectantly yielding an improved approximation of the one-time PDF, an approximate two-time PDF itself %
could be used directly to estimate various quantities of interest. An example is the distribution of the fluctuating heat in nonlinear, delayed systems~\cite{Sekimoto2010,Loos2019}. In future studies, we want to explore these possibilities with the new approximation discussed here.
\par
The development of further approximation schemes for the PDFs is beyond the scope of the present paper. However, besides the approach discussed in Sec.\,\ref{SEC:Approx}, another, quite different type of approximation is in fact readily given by the Markovian embedding approach, namely, by directly considering the extended Markovian system with a \textit{finite }number of auxiliary variables $n$~\cite{Niculescu2012}. Then, Eq.~(\ref{EQ:FPE-extended-2}) is a well-defined, closed FPE, which can be used to calculate the $n$-point PDF by means of standard techniques for Markovian FPEs (at least numerically). By marginalization of the resulting PDFs, one would recover the corresponding (approximate) PDFs of the delayed process. On the level of the stochastic delay differential equation (the LE), this is in fact a common method~\cite{Rosinberg2018,De1991,Rateitschak2007,Gupta2011}. It amounts to approximating the discrete delay by a Gamma-distributed memory kernel~(\ref{EQ:LE-kernel}) with finite width. However, a systematic investigation of this approximation (including an error estimate), has, to the best of our knowledge, not been performed yet. A similar finite-dimensional approximation for linear systems is discussed in~\cite{Rene2017}.
A more detailed study of these approximations is subject of ongoing work.
\par

Finally, we would like to reconsider the Markovian embedding from a conceptual point of view. Non-Markovianity, in general, and specifically the discrete time delay considered here, is a consequence of an incomplete description of a system, i.\,e., missing degrees of freedom. 
The delayed force might either stem from a specifically designed external agent, a \textit{feedback controller}, or, more generally, from any part of a more complex system (the ``super-system''), which is not explicitly modeled within the considered description, but interacts with the system at hand in form of delayed forces. The Markovian auxiliary system, in which we here embed the delayed process, allows us to mimic the external agent (or other parts of the  ``super-system'') without explicitly modeling it in detail. 
In this sense, the auxiliary variables can be interpreted as a ``substitute'' of the hidden degrees of freedom in the system. On the other hand, it is well-known that hidden degrees of freedom have a significant impact on the entropy production and other thermodynamical notions, see for example~\cite{Mehl2012,Kawaguchi2013}. Therefore, a study of the consequences of our treatment of delayed systems from a stochastic thermodynamical (and information-theoretical) perspective would be very interesting. This will be a focus of future work.}
\section*{Acknowledgements}
This work was funded by the Deutsche Forschungsgemeinschaft (DFG, German Research Foundation) - Projektnummer 163436311 - SFB 910. We further thank Christian Kuehn for fruitful discussions.
%
%
%
%
\appendix
%
%
%
%
%
%
\section{APPENDIX: The discrete delayed case as limit of the coarse-grained dynamics}\label{SEC:App1}
\small{
In this appendix, we show how the dynamical equation of the delayed system~(\ref{EQ:LE-orignial}) can be recovered by taking the limit $n\to\infty$ of the corresponding equation for the embedded system~(\ref{EQ:extended-LE}). To this end, we project onto the the dynamics of the variable $X_0$. Until completion of the projection, we will keep $n$ finite.
Formally integrating Eq.~(\ref{EQ:extended-LE}b) (by first solving the homogeneous problem and then finding a specific solution by variation of constants) yields
\begin{equation}\label{EQ:formal-sol-step1}
X_j(t)= X_j(-\tau) e^{-n (t+\tau) /\tau }+\frac{n}{\tau}\int_{-\tau}^{t} e^{-n(t-s)/\tau } X_{j-1}(s) \mathrm{d}s, ~~~\forall j \in \{1,2,...,n \}\, .
\end{equation}
%

In principle, this solution can be simplified using an appropriate initial condition for the auxiliary variables, for example $X_j(0) = 0, ~\forall j>0$.
However, we here proceed with the general case, i.\,e., without specifying the initial condition. 
%
%
As a next step (assuming $n>1$), we iteratively plug Eq.~(\ref{EQ:formal-sol-step1}) into the corresponding solution for $j+1$ [given by Eq.~(\ref{EQ:formal-sol-step1}) with $j+1$ instead of $j$]. For $1\leq j < n$, we find
\begin{subequations}
\begin{align}\label{EQ:Theta_Integrations}
X_{j+1}(t)& =  \left[ X_{j+1}(-\tau) +\frac{n}{\tau}  X_{j}(-\tau) \,t\right] e^{-n (t+\tau) /\tau }+\frac{n^2}{\tau^2} \int_{-\tau}^{t}  \left\{ \int_{-\tau}^{s} e^{-n (t-s')/\tau} X_{j-1}(s') \mathrm{d}s' \right\} \mathrm{d}s.
\end{align}
Using the Heaviside function defined by $\Theta(z)=0 ,\forall\, z < 0$ and $\Theta(z)=1,\forall\, 0\leq z$, we can simplify the double integral,
\begin{align}
X_{j+1}(t)
 = &  \left[  X_{j+1}(-\tau) +
 \frac{n}{\tau}  X_{j}(-\tau) \,t \right] e^{-n (t+\tau) /\tau } + \frac{n^2}{\tau^2} \int_{-\tau}^{s} \left\{\int_{-\tau}^{t}  \Theta (s-s') \mathrm{d}s \right\}  e^{-n (t-s')/\tau} X_{j-1}(s') \mathrm{d}s' , \\
 =&  \left[   X_{j+1}(-\tau) +
 \frac{n}{\tau}  X_{j}(-\tau) \,t \right] e^{-n (t+\tau) /\tau } + \int_{-\tau}^{t} \frac{n^2}{\tau^2} [t-s'] e^{-n (t-s')/\tau} X_{j-1}(s') \mathrm{d}s' . \label{EQ:formal-sol-step2}
\end{align}
Repeating this iterative procedure, we finally obtain for $j=n$
\begin{equation} \label{EQ:formal-sol-stepm}
X_{n}(t)= \left[X_{n}(-\tau)+\frac{n}{\tau}X_{n-1}(-\tau)t+\frac{n^2}{\tau^2}X_{n-2}(-\tau)t^2+...\right]\, e^{-n (t+\tau) /\tau } + \int_{-\tau}^{t} K_n(t-s) X_{0}(s) \mathrm{d}s ,
\end{equation}
\end{subequations}
with the Gamma-distributed \textit{memory kernel} defined in Eq.~(\ref{EQ:LE-kernel}). The first term in~(\ref{EQ:formal-sol-stepm}) vanishes trivially, if the auxiliary variables satisfy the initial condition $X_{j\in\{1,2,..,n\}}(-\tau)=0$.
Interestingly, this terms also vanishes if the asymptotic behavior at $t\to\infty$ is considered, due to the exponential damping with time. 
Plugging~(\ref{EQ:formal-sol-stepm}) into (\ref{EQ:extended-LE}a) [neglecting the initial condition term] then yields the projected Eq.~(\ref{EQ:LE-kernel}). 
\par
Now we consider the limit $n\to\infty$ of the memory kernel~(\ref{EQ:LE-kernel}). First, we find that the mean value of the kernel and the integral over it are, independently of the value of $n$, given by $\mu =\int_{0}^{\infty} K_n(T)T \mathrm{d}T =\tau$ and $\int_{0}^{\infty} K_n(T) \mathrm{d}T =1$, respectively. Furthermore, for large $n$, the value of the kernel at $\tau$ can be estimated by using the Stirling formula $n!\approx \sqrt{2\pi n}(n/e)^n$, yielding $K_n(\tau)= \frac{{n}^n \,e^{-n}}{(n-1)! \tau} \approx\frac{n!}{(n-1)! \sqrt{2\pi n}\tau}=\sqrt{\frac{n }{ 2\pi}}\frac{1}{\tau}\to \infty.$ 
On the other hand, the variance of the kernel vanishes for $n\to \infty$, since $\int_{0}^{\infty} K_n(T)T^2 \mathrm{d}T -\mu^2=\tau^2/{n}\to 0$. This implies %
\begin{equation}
\lim_{n  \rightarrow \infty} K_n(T)= \delta(T-\tau).
\end{equation}
Thus, the projected equation~(\ref{EQ:LE-kernel}) is indeed equal to Eq.~(\ref{EQ:LE-orignial}) in the limit $n\to \infty$. We note that even the transient dynamics (i.\,e., finite $t$) is only equivalent for the initial conditions $X_{j\in\{1,2,..,n\}}(-\tau)\equiv 0$, as mentioned above.
}
\section{APPENDIX: Connection to Fokker-Planck hierarchy from Novikov's theorem}\label{APP:Nov-Embedd-Equivalence}
\small{
While the FPE for $\rho_1$ obtained by our approach is equivalent to the first member of the FP hierarchy from Novikov's theorem~(\ref{EQ:dFPE_start}), the second members differ. However, the apparent disagreement can be resolved. Indeed, as we show below, our result~(\ref{EQ:dFPE_2nd-member_2}) can be transformed to the representation obtained from Novikov's theorem [given in~(\ref{EQ:General_dFPE2})]. We focus on the term which differs, i.\,e., the drift term $\partial_{x_\tau} [\rho_2(x,t;x_\tau,t-\tau)$ $ \langle \dot{X} ( t - \tau ) |x=X(t) , x_\tau=X({t - \tau})\rangle]$.
To this end, we first use the definition of $\rho_2$ via delta-distributions, plug in the LE~(\ref{EQ:LE-orignial}), which yields 
\begin{subequations}
\begin{align}
 \partial_{x_\tau} \left[\langle \dot{X} ( t - \tau ) |x=X(t) , x_\tau=X({t - \tau})\rangle \rho_2(x,t;x_\tau,t-\tau)  \right]
 %
%
=&~
\partial_{x_\tau} \left\langle \dot{X}(t-\tau) \,\delta[X(t)-x]\delta[X(t-\tau)-x_\tau] \right\rangle 
\nn
\stackrel{\mathrm{LE}}{=}&
\partial_{x_\tau} \left\langle F[X(t-\tau),X(t-2\tau)]\,\delta[X(t)-x]\delta[X(t-\tau)-x_\tau] \right\rangle 
\nn
&+
\partial_{x_\tau} \left\langle \sqrt{2D_0}\xi(t-\tau) \,\delta[X(t)-x]\delta[X(t-\tau)-x_\tau] \right\rangle 
.
\label{Zw1}
\end{align} 
At this point, we have two correlations to deal with. The first one (involving $F_{}$) can easily be simplified by expressing the ensemble average with the help of the PDF $\rho_3$, i.~e.,
\begin{align}\label{Zw1b}
\partial_{x_\tau} \Big\langle  F\left[X(t-\tau),X(t-2\tau)\right]& \,\delta[X(t)-x]\delta[X(t-\tau)-x_\tau] \Big\rangle \nn
=&\partial_{x_\tau} \iiint_{\Omega }^{ }F_{}(\xt,\xtt) \,\delta[X(t)-x]\delta[X(t-\tau)-x_\tau]\rho_3(x,t;x_\tau,t-\tau;x_{2\tau},t-2\tau) \mathrm{d}x \mathrm{d}x_\tau \mathrm{d}x_{2\tau}
\nn
=&\partial_{x_\tau} \int_{\Omega }^{ } F_{}(\xt,\xtt) \rho_3(x,t;x_\tau,t-\tau;x_{2\tau},t-2\tau) \mathrm{d}x_{2\tau}.
\end{align}
The remaining term on the right side of~(\ref{Zw1}) (involving $\xi$) is simplified with the help of Novikov's theorem~(\ref{NOV-THE}) (see Appendix~\ref{APP:Nov-Theorem}). Specifically, we define $\Lambda[\xi]:= \delta[x-X(t)]\, \delta[x_\tau-X(t\!-\!\tau)] $, which yields
\begin{align}\label{Zw2}
\partial_{x_\tau} \left\langle \sqrt{2D_0}\xi(t-\tau) \,\delta[X(t)-x]\delta[X(t-\tau)-x_\tau] \right\rangle=&\sqrt{2D_0}\partial_{x_\tau} \left\langle \xi(t-\tau) \Lambda[\xi] \right\rangle, \nn
=&\sqrt{2D_0} \partial_{x_\tau} \left\langle \frac{\delta  \Lambda[\xi]  }{\delta \xi(t-\tau)} \right\rangle \nn
=&\sqrt{2D_0}\partial_{x_\tau} \left[ \left\langle \frac{\delta  \Lambda[\xi]  }{\delta X(t)}\frac{\delta  X(t)  }{\delta \xi(t-\tau)} \right\rangle+\left\langle \frac{\delta  \Lambda[\xi]  }{\delta X(t-\tau)} \frac{\delta  X(t-\tau)  }{\delta \xi(t-\tau)}\right\rangle \right]
\nn
=&-\sqrt{2D_0}\partial_{x_\tau} \left\langle \partial_{x }\delta[x-X(t)]\, \delta[x_\tau-X(t\!-\!\tau)] {\frac{\delta  X(t)  }{\delta \xi(t-\tau)}} \right\rangle 
\nn
&-\sqrt{2D_0}\partial_{x_\tau}\left\langle \partial_{x_\tau} \delta[x-X(t)]\, \delta[x_\tau-X(t\!-\!\tau)]\underbrace{\frac{\delta  X(t-\tau)  }{\delta \xi(t-\tau)}}_{\psi}\right\rangle .
\end{align}
In the last step we have plugged in the definition of $\Lambda$ and used the chain rule. Now, the term denoted $\psi$ is calculated by replacing $X(t-\tau)$ with the integral form of the LE~(\ref{EQ:LE-orignial}), i.\,e., $X(t')= \int_{0}^{t'}\left[ F[X(s),X(s-\tau)]+\sqrt{2D_0}\xi(s)\right]\mathrm{d}s$, $\forall\,t'>0$, yielding
\begin{align}
\psi=\frac{\delta X(t-\tau)}{\delta \xi(t-\tau)} =& \frac{\delta }{\delta \xi(t-\tau)} \int_{0}^{t-\tau} \left[  F[X(s),X(s-\tau)]+\sqrt{2D_0}\xi(s)\right]\mathrm{d}s 
=\sqrt{2D_0}\,\int_{0}^{t-\tau} \delta[s-(t-\tau)]\mathrm{d}s=\sqrt{\frac{D_0}{2}},
\end{align}
independently from the specific form of $F_{}$.
We note that the other functional derivative in~(\ref{Zw2}) cannot be treated analogously, since evaluation of the functional derivative of the trajectory $X$ w.\,r.\,t. the earlier noise $\xi$ requires the formal solution of~(\ref{EQ:LE-orignial}), which is only known for linear systems. Indeed, for a LE with $ F= -c_1 X(t) -c_2 X_{}(t-\tau)$, one finds $\frac{\delta  X(t)  }{\delta \xi(t-\tau)}=\sqrt{2D_0} e^{-c_1 \tau}$ by using the method of steps, see~\cite{Loos2017}.
For general nonlinear forces $F$, this term must be treated differently, e.\,g. by approximation methods, but a discussion of this goes beyond the scope of this paper. We therefore keep this functional derivative and only evaluate the delta-distributions. Based on these considerations, we can rewrite~(\ref{Zw2}) as
\begin{align}
\sqrt{2D_0}\,\partial_{x_\tau} \left\langle \xi(t-\tau) \Lambda[\xi] \right\rangle=-\sqrt{2D_0}\,\partial_{x_\tau}\partial_{x } \left\langle \frac{\delta  X(t)  }{\delta \xi(t-\tau)}\Big|_{\!\!\!\!\!\!\!\!\!\!  X(t)=x \atop\!\!  X(t - \tau)=x_\tau} \right\rangle \rho_2(x,t;x_{\tau},t-\tau) -D_0\,\partial_{x_\tau}^2\rho_2(x,t;x_{\tau},t-\tau).
\end{align}
\end{subequations}
In combination with (\ref{Zw1},\ref{Zw1b}), we obtain the identity
\begin{align}
\partial_{x_\tau} \left\langle \dot{X}(t-\tau) \,\delta[X(t)-x]\delta[X(t-\tau)-x_\tau] \right\rangle 
=&
-\sqrt{2D_0}\partial_{x_\tau}\partial_{x}   \left\langle \frac{\delta  X(t)  }{\delta \xi(t-\tau)}\Big|_{\!\!\!\!\!\!\!\! \!\!   X(t)=x \atop\!\!   X(t - \tau)=x_\tau} \right\rangle \rho_2(x,t;x_{\tau},t-\tau) 
\nn
&-D_0 \partial_{x_\tau}^2 \rho_2(x,t;x_{\tau},t-\tau)+\partial_{x_\tau}  \int_{\Omega }^{ } F_{}(\xt,\xtt) \rho_3(x,t;x_\tau,t-\tau;x_{2\tau},t-2\tau) \mathrm{d}x_{2\tau}
\nn
=&
\partial_{x_\tau} \left[\langle \dot{X} ( t - \tau ) |x=X(t) , x_\tau=X({t - \tau})\rangle \rho_2(x,t;x_\tau,t-\tau)  \right]
,
\end{align}
proving that the second member of the here presented FP hierarchy, Eq.~(\ref{EQ:dFPE_2nd-member_2}), is identical to the corresponding one~(\ref{EQ:General_dFPE2}) obtained from Novikov's theorem.
}
\section{APPENDIX: Novikov's theorem}\label{APP:Nov-Theorem}
\small{
In this section, we establish the relation~\cite{Novikov1965}
 \begin{equation}\label{NOV-THE}
\big \langle\,\Lambda[\xi]  \xi(t) \,\big \rangle = \Bigg \langle \frac{\delta \Lambda[\xi]}{\delta \xi(t)}\Bigg\rangle,
\end{equation}
for a functional $\Lambda$ of a \textit{Gaussian white noise} $\xi$. It links the functional derivative w.\,r.\,t. the noise to the cross-correlation between functional and noise. First, we consider the case where the ensemble is w.\,r.\,t. a \textit{fixed }initial condition $\phi=X(-\tau\leq t\leq t)$, yielding $\Lambda(0)={\Lambda_0}$. In a second step we will generalize towards initial conditions drawn from an arbitrary distribution. 

\begin{subequations}
We express the ensemble average $\langle ..\rangle_{\Lambda_0}$ of the left hand side of~(\ref{NOV-THE}) via the path integral over all possible ``paths'' $\xi$ between time $0$ and $t$, accounting for all possible realizations of the random process $\xi$ at each instant in time
\begin{align}
\left \langle\Lambda[\xi]\xi(t) \right \rangle_{\Lambda_0} =  \int_{\xi_0}^{\xi_t}\Lambda\left[\xi|{\Lambda_0}\right]\,\xi \, \mathcal{P}[\xi]\, \mathcal{D}[\xi], \label{EQ:Nov-step0}
\end{align}
with arbitrary but fixed $\xi_0:=\xi(t_0)$ and $\xi_t:=\xi(t)$. Please note that specifying the noise process at the boundaries does not impose a restriction on the generality as we deal with \textit{white} noise. The weight $\mathcal{P}[\xi]$ of each white noise realization is given by the Gaussian path probability (for arbitrary $t>0$)
\begin{align}
\mathcal{P}[\xi]=\mathcal{J} e^{-\frac{1}{2}\int_{0}^{t}\xi({t'})^2 \mathrm d t'},
\end{align}
with Jacobian $\mathcal{J}$.
Now, we rewrite the integrand in (\ref{EQ:Nov-step0}) using that the functional derivative of $\mathcal{P}[\xi]$ w.\,r.\,t. $ \xi$ is simply $(-\xi) \mathcal{P}[\xi]$, yielding
\begin{align}
\left \langle\Lambda[\xi]\xi(t) \right \rangle_{\Lambda_0}  = & - \int_{\xi_0}^{\xi_t}\Lambda\left[\xi|{\Lambda_0}\right] \frac{\delta}{\delta \xi}\!\left\{ e^{-\frac{1}{2}\int_{0}^{t}\xi({t'})^2 \mathrm d t'}\right\} \,\mathcal{D}[\xi] \nonumber\\
 = & - \int_{\xi_0}^{\xi_t} \frac{\delta\left\{ \Lambda[\xi]  \mathcal{P}[\xi] \right\}}{\delta \xi} \,\mathcal{D}[\xi] + \int_{\xi_0}^{\xi_t}\frac{\delta\Lambda[\xi]}{\delta \xi}  \mathcal{P}[\xi] \,\mathcal{D}[\xi]. \label{EQ:Nov-step1}
\end{align}
In the last step we have used the product rule (and omitted the initial condition $\Lambda_0$ for sake of a shorter notation).
The functional derivative in the first path integral yields
\begin{align}
\int_{\xi_0}^{\xi_t}\frac{\delta\left\{ \Lambda[\xi]  \mathcal{P}[\xi] \right\}}{\delta \xi} \mathcal{D}[\xi]=
\int_{\xi_0}^{\xi_t} \Lambda[\xi +\delta \xi]  \mathcal{P}[\xi+\delta \xi]  \mathcal{D}[\xi]-\int_{\xi_0}^{\xi_t}\Lambda[\xi]  \mathcal{P}[\xi]  \mathcal{D}[\xi].
\end{align}
Now we use that the variations of the paths $\xi+\delta \xi$ are already contained in the integral over \textit{all} paths, and find 
\begin{align}
\int_{\xi_0}^{\xi_t}\frac{\delta\left\{ \Lambda[\xi]  \mathcal{P}[\xi] \right\}}{\delta \xi} \mathcal{D}[\xi]=
\int_{\xi_0}^{\xi_t}\Lambda[\xi]  \mathcal{P}[\xi]  \mathcal{D}[\xi]-\int_{\xi_0}^{\xi_t}\Lambda[\xi]  \mathcal{P}[\xi]  \mathcal{D}[\xi]=0,
\end{align}
(as long as $\int \Lambda[\xi] \mathcal{P}[\xi] \mathcal{D}[\xi] <\infty $),
see also~\cite{Hall2016} (on p.\,273f). 
Hence, we obtain from (\ref{EQ:Nov-step1}),
\begin{align}
\left \langle\Lambda[\xi]\xi(t) \right \rangle_{\Lambda_0} = & \int_{\xi_0}^{\xi_t}\frac{\delta\Lambda[\xi]}{\delta \xi}  \mathcal{P}[\xi] \,\mathcal{D}[\xi] = \left\langle \frac{\delta\Lambda[\xi]}{\delta \xi}  \right\rangle_{\Lambda_0}. \label{EQ:Nov-step2}
\end{align}
\end{subequations}
\par
This result can readily be generalized to ensembles where the initial conditions are instead drawn from an arbitrary, normalized distribution $P(\Lambda_0)$. In particular,
Eq.~(\ref{EQ:Nov-step2}) implies
  \begin{align} 
\left \langle\Lambda[\xi]\xi(t) \right \rangle = \int \left \langle\Lambda[\xi]\xi(t) \right\rangle_{\Lambda_0} P(\Lambda_0) \mathrm{d}\Lambda_0 = 
\int \left \langle \frac{\delta\Lambda[\xi]}{\delta \xi}  \right \rangle_{\Lambda_0} P(\Lambda_0) \mathrm{d}\Lambda_0 
=\left\langle \frac{\delta\Lambda[\xi]}{\delta \xi}  \right\rangle.
\end{align}
This is the relation (\ref{NOV-THE}), which is often referred to as \textit{Novikov's theorem}.
}

\section{APPENDIX: The approximation in the regime of short delay times}\label{APP:Short-tau}
\begin{figure}
\begin{minipage}{0.45\textwidth}
\large{(a) $k=2$, $\tau=0.1$}\\
\includegraphics[width=.85\textwidth]{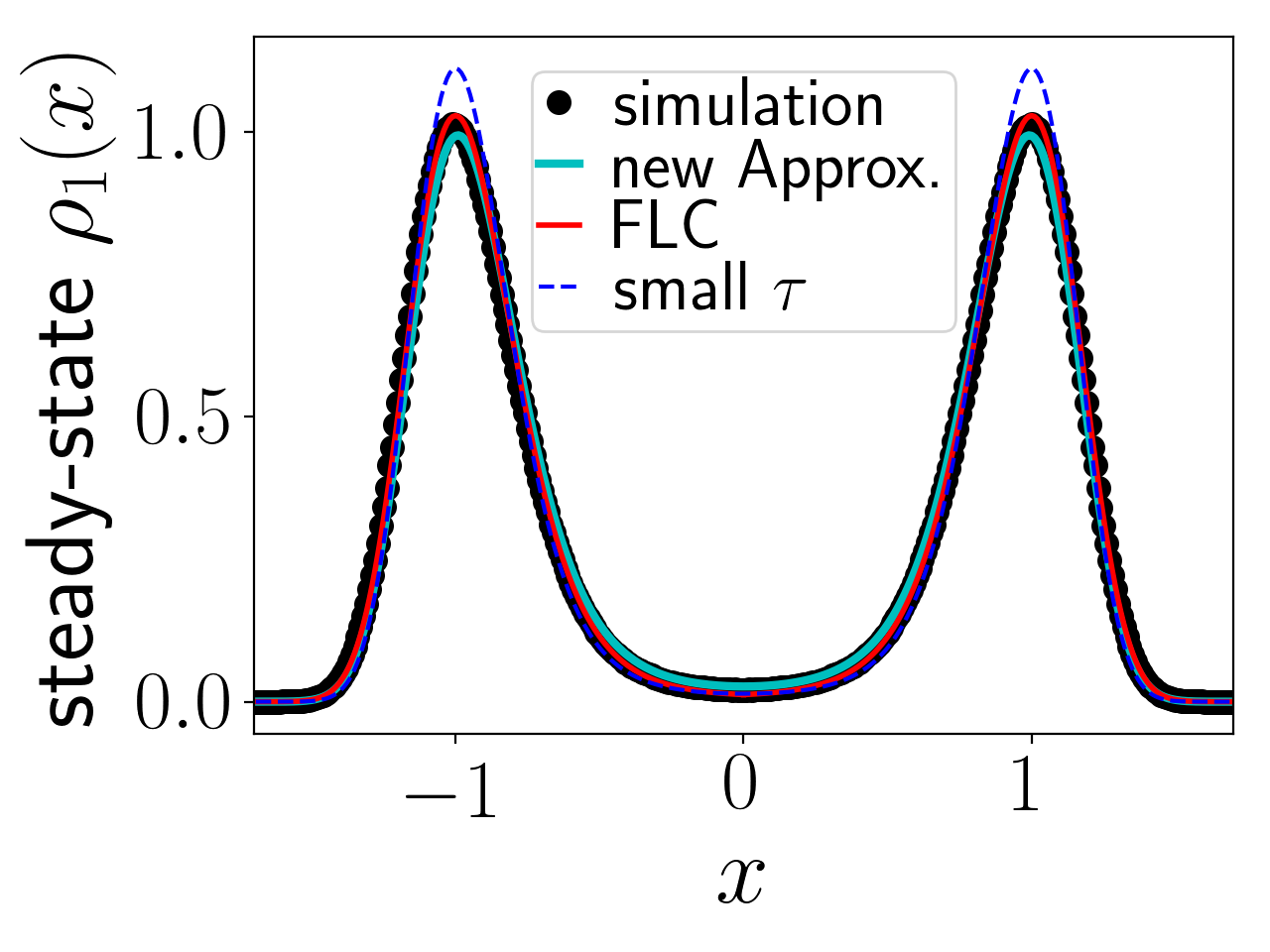}
\end{minipage}
\hfill
\begin{minipage}{0.45\textwidth}
\large{(b) $k=-10$, $\tau=0.1$}\\
\includegraphics[width=.85\textwidth]{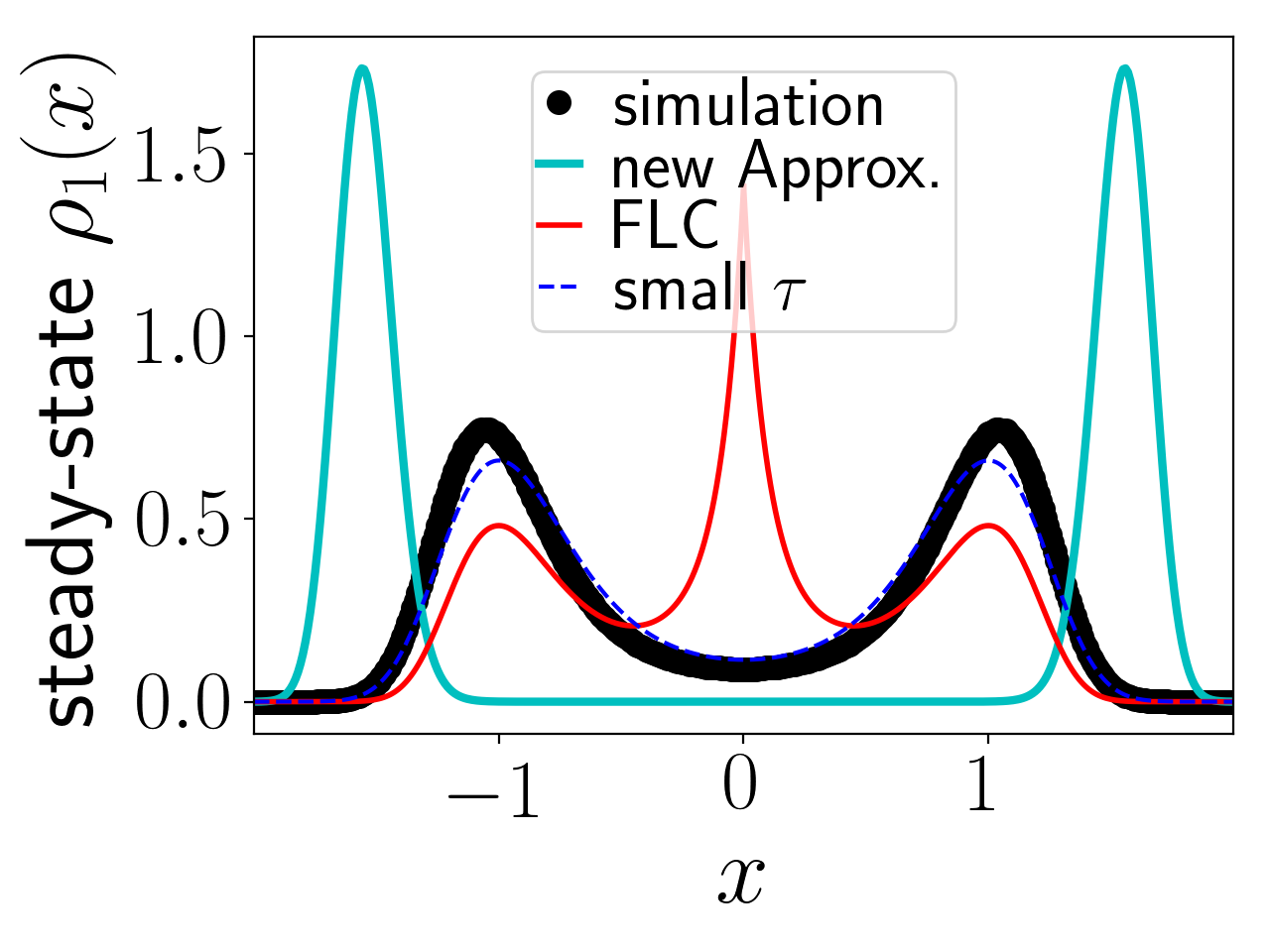}
\end{minipage}
\caption
{
Comparison of different approximations for the one-time PDF in the doublewell potential with linear delay force $k X(t-\tau)$ at $V_0=3.5$, for two exemplary values of $k$. In contrast to Fig.\,\ref{FIG:5} where $\tau=10$, here a \textit{shorter} delay time of $\tau=0.1$ is considered.
}\label{FIG:6}
\end{figure}

\red{The approximate two-time PDF~(\ref{EQ:Approx_bistable}) is, in general, not reliable, if the delay time is very short, as the underlying assumption is violated. For the moderate $|k|/V_0$ and small $\tau$ values considered in Fig.\,\ref{FIG:6}\,(a), the corresponding one-time PDF (obtained by marginalization) still gives a reasonable approximation. Here also the other approaches from the literature are justified. For the large (and negative) $|k|/V_0$ value in Fig.\,\ref{FIG:6}\,(b), only the small $\tau$ expansion yields a reasonable approximation.}

\section*{References}

\renewcommand\refname{ }
\renewcommand\refname{\vskip -1cm}
\bibliographystyle{abbrv}
\setlength{\bibsep}{0pt plus 0.3ex}



\end{document}